\documentclass[twocolumn]{aastex63}

\usepackage{float}
\usepackage{amsmath,amssymb}
\usepackage{natbib}    % For use with bibtex
\usepackage{hyperref} % For internal hyperlinks
\usepackage{graphicx} % For included graphics
\usepackage{color}
\usepackage{verbatim}    
\usepackage{enumitem}
\usepackage{natbib}
\usepackage{CJK}
\usepackage{multirow}

\newcommand{\proptosim}{\mathrel{\vcenter{
 \offinterlineskip\halign{\hfil$##$\cr
 \propto\cr\noalign{\kern2pt}\sim\cr\noalign{\kern-2pt}}}}}
\renewcommand{\b}[1]{\boldsymbol{#1}}

\newcommand{\mean}[1]{\langle #1\rangle}

\newcommand{\au}{\mathrm{AU}}
\newcommand{\cm}{\mathrm{cm}}
\newcommand{\m}{\mathrm{m}}

\newcommand{\g}{\mathrm{g}}
\newcommand{\G}{\mathrm{G}}
\newcommand{\K}{\mathrm{K}}  
\newcommand{\km}{\mathrm{km}}
\newcommand{\kms}{\mathrm{km\ s^{-1}}}

\newcommand{\eV}{\mathrm{eV}}
\newcommand{\keV}{\mathrm{keV}}
\newcommand{\s}{\mathrm{s}}
\newcommand{\yr}{\mathrm{yr}}
\newcommand{\hr}{\mathrm{hr}}
\newcommand{\ang}{\ensuremath{\text{\AA}}}
\newcommand{\dyn}{\mathrm{dyn}}
\newcommand{\lya}{\text{Ly}\ensuremath{\alpha~}}

\newcommand{\B}{\mathrm{B}}
\renewcommand{\d}{\mathrm{d}}
\newcommand{\e}{\mathrm{e}}

\renewcommand{\ion}[2]
  {{\rm#1}\;\textsc{\MakeLowercase{#2}}}

\newcommand{\eff}{\mathrm{eff}}

\newcommand{\wind}{\mathrm{wind}}

\newcommand{\p}{{\rm p}}
\newcommand{\J}{{\rm Jup}}

\newcommand{\eq}{\mathrm{eq}}   % equal-temperature
\newcommand{\rcb}{\mathrm{rcb}} % radiative-convective
\newcommand*\chem[1]{\ensuremath{\mathrm{#1}}}
\newcommand{\pos}[1]{\ensuremath{\mathrm{#1}^+}}

\newcommand{\ext}[1]{\ensuremath{\mathrm{#1}^*}}

\newcommand{\figdir}{Figures}
\renewcommand{\roman}[1]{\textsc{\MakeLowercase{#1}}} 

\newcommand{\HeIline}{\ion{He}{I} $10830~\ang$\ }
\begin{document} 

\title{Metastable Helium
  Absorptions with 3D Hydrodynamics and \\
  Self-Consistent Photochemistry II: \\
  WASP-107\lowercase{b}, Stellar
  Wind, Radiation Pressure, and Shear Instability}

\author{Lile Wang$^{1}$, Fei Dai$^{2}$}

\footnotetext[1]{Center for Computational Astrophysics,
  Flatiron Institute, \\ New York, NY 10010;
  lwang@flatironinstitute.org}
\footnotetext[2]{Division of
  Geological and Planetary Sciences, \\
  California Institute of
  Technology, Pasadena, CA 91125} 

\begin{abstract}
  This paper presents simulations of the metastable helium
  (\ext{He}) observations of WASP-107b, so far the highest
  signal-to-noise ratio detection that is confirmed by three
  different instruments. We employ full 3D hydrodynamics
  coupled with co-evolving non-equilibrium thermochemistry
  and ray-tracing radiation, predicting mass loss rates,
  temperature profiles, and synthetic \ext{He} line profiles
  and light curves from first principles. We found that a
  stellar wind stronger than solar is demanded by the
  observed heavily blueshifted line profile and asymmetric
  transit light curve. Contrary to previous beliefs, we
  argue that radiation pressure can be important for \lya
  observations but {\it not} \ext{He}. We found WASP-107b is
  losing mass at a rate of
  $\dot{M} \simeq 1.0\times
  10^{-9}~M_\oplus~\yr^{-1}$. Although $\dot{M}$ varies by
  $<1~\%$ given constant wind and irradiation from the host,
  shear instabilities still emerge from wind impacts,
  producing $\sim 10~\%$ fluctuations of \ext{He} transit
  depths over hour-long timescales. The common assumption
  that \ext{He} transit depth indicates the fluctuation of
  $\dot{M}$ is problematic. The trailing tail is more
  susceptible than planet adjacency to the shear
  instabilities, thus the line profile is more variable in
  the blue-shifted wing, while the transit light curve is
  more variable after mid-transit. We stress the synergy
  between \lya (higher altitudes, lower density) and
  \ext{He} (lower altitudes, higher density) transit
  observations, particularly simultaneous ones, yield better
  understanding of planetary outflows and stellar wind
  properties.
\end{abstract}

\keywords{planets and satellites: atmospheres --- planets
  and satellites: composition --- planets and satellites:
  physical evolution --- method: numerical }

\section{Introduction}
\label{sec:intro}

The ``\HeIline line'' or the ``metastable helium line''
(\ext{He} line for short) transitions between the $2^3$S and
the upper $2^3$P$_J$ ($J=0, 1, 2$) states are radiatively
decoupled from the ground state (for a magnetic dipole
transition) and have slow spontaneous decay rates:
$A \simeq 1.3\times 10^{-3}~\s^{-1}$
\citep{1971PhRvA...3..908D}. The high cosmic abundance of
helium, the absence of interstellar absorption and the
observability from the ground together enable the \ext{He}
lines as a unique probe of atmospheric outflows from
exoplanets. The first secure detection of \ext{He} in
transmission was made for WASP-107b with the Hubble Space
Telescope \citep{2018Natur.557...68S}. This detection
rekindled decade-long interest in this transition
\citep{2000ApJ...537..916S, 2016MNRAS.458.3880T,
  2018ApJ...855L..11O}; many more detections around other
exoplanets have been made since then \citep[e.g.][]
{2018Sci...362.1384A, 2018Sci...362.1388N,
  2018A&A...620A..97S, 2020AJ....159..115K,
  2020ApJ...894...97N}.

There is no surprise that WASP-107b was the first exoplanet
to show \ext{He} detection. The host is young
($\sim 600~{\rm Myr}$ from gyrochronology) and active
(chromospheric activity index $S = 0.89$), expected to give
out strong high-energy radiation that powers planetary
photoevaporative outflow. The host's spectral type is K6,
which is right at the sweet spot of EUV-FUV flux ratio that
maximally favors the \ext{He} absorption
(\citealt{2019ApJ...881..133O}; Paper I). The planet is
puffy and susceptible to outflows: it has a mass of an icy
giant ($0.12~M_{\rm Jup}$) but a radius closer to that of
Jupiter ($0.94~M_{\rm Jup}$). The optical transit depth is
$\sim 2.2~\%$, while the \ext{He} transit depth is a
whopping $\sim 7~\%$. The mean density of the planet is only
$\sim 0.19~\g~\cm^{-3}$, which is reminiscent of the
anomalously low-density planets "super-puffs"
\citep[e.g.][]{2020AJ....159...57L,
  2020AJ....160..201C}. \citet{2019ApJ...873L...1W} and
\citet{2020ApJ...890...93G} proposed that ``super-puffs''
may appear inflated due to high-altitude dusts or hazes
elevated by an outflowing atmosphere; the near-infrared HST
observation \citep{2018ApJ...858L...6K} is indeed suggestive
of high-altitude condensates partially muting the
transmission features. WASP-107b is also dynamically
interesting. \citet{2017AJ....153..205D} suspected the
planet is on a polar orbit due to the lack of repeating
spot-crossing events. More recently, a Rossiter-McLaughlin
measurement by Rubenzahl, R. et al (in prep) confirmed this
suspicion and thus demands a dynamically hot formation and
evolution pathway that may involve the non-transiting planet
107c ($P\sim 1100~{\rm days}$, $M\sin i\sim 110~M_\oplus$;
Piaulet, C. et al., submitted)

These reasons render WASP-107b a unique and interesting
system. The \ext{He} transits of WASP-107b have been
observed multiple times \citep{2018Natur.557...68S,
  2018Sci...362.1384A, 2020AJ....159..115K}. Its
well-resolved line profile exhibits asymmetric shape skewing
towards the blue-shifted wing, which suggests a comet-like
tail. The line ratio of three different $2^3$P$_J$ levels
also seems to deviate from the simple quantum degeneracy
ratio of 1:3:5 (or 1:8 given that the two latters are not
distinguishable). Having been confirmed by different
instruments (CARMENES; Keck/NIRSPEC), these features are
likely attributed to the morphology and kinematics of
planetary outflows, rather than instrumental effects.

The \ext{He} dataset of WASP-107b potentially very revealing
and should be analyzed in detail. Two models have been
presented in the literature: the 1D isothermal model by
\citet{2018ApJ...855L..11O} assumes the density and velocity
profile of a Parker Wind \citep{Parker}. The model has to
assume, rather than predict, the mass loss rate and
temperature of the outflow. Moreover, 1D models are unable
to capture the full orbital dynamics and cannot produce a
comet-like tail. The alternative model is the EVaporating
Exoplanets code \citep[EVE; see][]{2015A&A...582A..65B,
  2018Sci...362.1384A}. The lower layer (thermosphere) of
this model is also a Parker Wind solution; the upper layer
is a Monte-Carlo particle simulations with He particles
under the influence of planetary and stellar gravity and
radiation pressure. As we will elaborate later in this work,
nevertheless, particle-based code may not be appropriate for
simulating \ext{He} transits. The authors of EVE
acknowledged that the particle-based treatment of radiation
pressure accelerates the outflow way too quickly: they had
to artificially decrease the stellar spectrum near the
\ext{He} lines by a factor of $\sim 50$ to achieve
reasonable agreement with the observations of WASP-107b.

In this work, we applied to WASP-107b our model that
conducts 3D hydrodynamics, self-consistent thermochemistry,
ray-tracing radiative transfer, and especially the processes
that populate and destroy \ext{He} (see the first paper in
the series, Wang, L. \& Dai, F., submitted; Paper I
hereafter). Starting from the observed stellar and planetary
properties, and making assumptions of the high energy
spectral energy distribution (SED) of the host, we can
predict the mass loss rate, the temperature profile, the
ionization states and the various \ext{He} simulations of
WASP-107b. We will also use WASP-107b as a case study to
investigate how stellar wind, radiation pressure and
shear instability affect planetary outflows and
their observability.

This paper is structured as follows: in \S 2, we will
briefly described our model and simulation setup. In \S 3,
we present the fiducial model of WASP-107b that shows
remarkable agreement with observation. \S 4, we perturb the
fiducial model in various parameters to investigate
\S\ref{sec:summary} summarize the paper and figure out
prospective improvements for the future.

\begin{figure*}
  \centering
  \hspace*{-0.1in} 
  \includegraphics[width=7.2in, keepaspectratio]
  {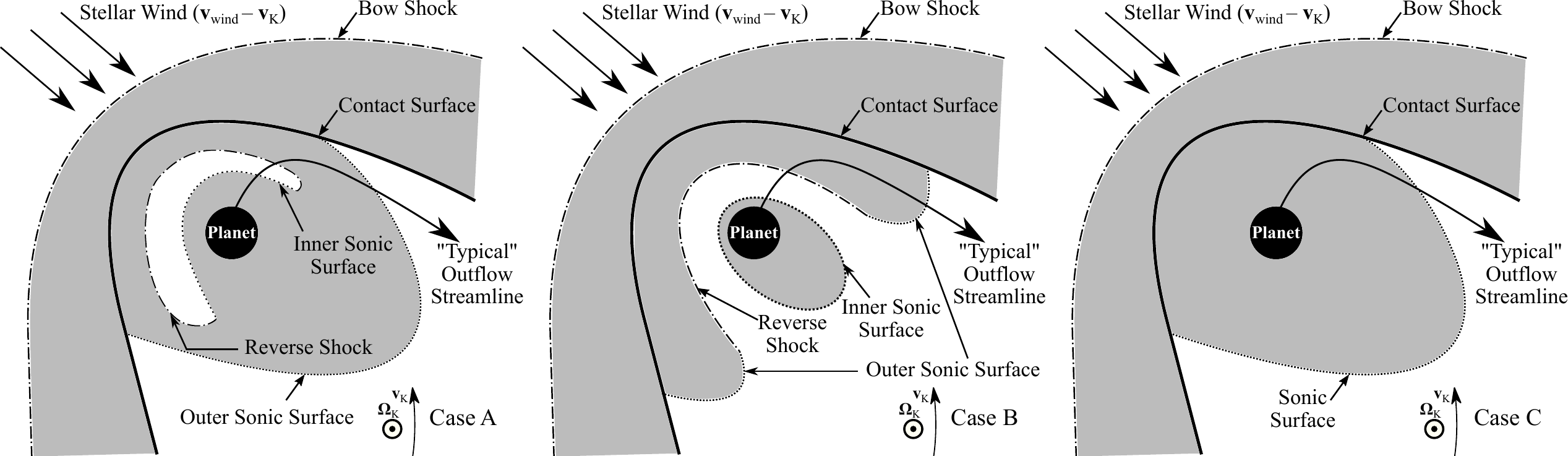}
  \caption{Schematic diagrams (not to scale) of the
    interactions between a stellar wind and a planetary
    atmospheric outflow (see also
    \S\ref{sec:method-wind}). In all panels, the host star
    resides to the left of the diagrams, and the planet's
    linear Keplerian velocity points upwards within the
    paper plane (see the indicators of $\mathbf{v}_\K$ and
    $\mathbf{\Omega}_\K$ at the lower-left corner of all
    panels). Subsonic regions are illustrated by grey
    shades. Typical streamlines are indicated by thin solid
    lines with arrow heads. Shock fronts, including the bow
    shocks and the reverse shock, are indicated by
    dash-dotted lines. Contact surfaces (contact
    discontinuities) are shown by heavy solid lines. Dotted
    lines indicate the sonic critical surfaces. {\bf The
      left panel}, marked as {\bf Case A}, shows a typical
    case that part of the the planet outflow experiences
    initial thermally driven acceleration, reverse shock
    deceleration (due to the stagnation and deflection by
    the stellar wind), and final acceleration by
    expansion. Note that the outer sonic surface is
    connected to the reverse shock, and any streamline
    crossing the outer sonic surface {\it must} go through
    the reverse shock front. {\bf The middle panel} presents
    {\bf Case B}, which is qualitatively similar to Case A
    but has different topologies. {\bf The right panel}
    presents another typical {\bf Case C}, which does not
    have a reverse shock: the planet outflow stays subsonic
    before the stagnation and deflection, and only become
    supersonic after it already escapes from the ``trap''
    shaped by the stellar wind. }
  \label{fig:wasp107b-schematic} 
\end{figure*}

\section{Methods}
\label{sec:method}

\subsection{Basic Setup}
\label{sec:method-setup}

Our numerical simulation suite was described in detail in
Paper I. For a brief re-cap, our simulation is carried in 3D
with gravity of both the host star and the planet, and the
effects of orbital motion including centrifugal and Coriolis
forces. Our code computes ray-tracing radiative transfer and
non-equilibrium thermochemistry simultaneously with
hydrodynamics. For simplicity, we assume circular orbit and
that the planet is tidally locked, adopting a co-rotating
planet-centric frame. We focus on the upper layer of the
atmosphere, including a quasi-isothermal layer assumed to
have equilibrium temperature, and an outflowing region
irradiated by high-energy photons. The boundary conditions
are set so that the observed mass and transit radius of the
planet can be reproduced. We describe the high-energy
spectral energy distribution (SED) of the host star with 6
characteristic energy bins: (1) $2~\eV$ for the
near-infrared and optical bands; (2) $7~\eV$ for far
ultraviolet (FUV) photons (``soft FUV''; note that these
photons can ionize \ext{He}); (3) $12~\eV$ for the
Lyman-Werner (``LW'') band FUV photons that photodissociate
molecular hydrogen, (4) $20~\eV$ for ``soft'' extreme
ultraviolet (``soft EUV'') photons that ionize hydrogen but
not helium, (5) $40~\eV$ for EUV and soft X-ray (``hard
EUV'') photons that ionize helium, and (6) $3~\keV$ photons
for the X-ray. We make synthetic observations including
\ext{He} line profiles and transit light curves in the
vicinity of \ext{He} line. To further facilitate comparisons
between observation and simulations, we compute summary
statistics such as the equivalent widths
$\mean{W_\lambda} \equiv \int\Delta
\epsilon(\lambda)~\d\lambda$, the radial velocity shift of
the absorption peak $\Delta v_{\rm peak}$, and the
full-width-half-maximum (FWHM) of the absorption line
profile.

Specifically for WASP-107, \citet{2017A&A...604A.110A}
reported a K6 host star with a mass of $M_* = 0.69~M_\odot$
and a radius of $R_* = 0.66~R_\odot$, and an effective
temperature of $T_{*,\eff} = 4430~\K$. Planet b orbits its
host on a near-circular but polar orbit \citep[][Piaulet,
C. et al, submitted; Rubenzahl, R. et al,
submitted]{2017AJ....153..205D}. The semi-major axis is
$a = 0.055~\au$ where the equilibrium temperature is
$T_\eq = 740~\K$; the transit light curve indicates a small
impact parameter \citep[$b=0.07$][]
{2017AJ....153..205D}. WASP-107b has an optical transiting
radius of $R_\p \simeq 0.94~R_\J$ and a mass
$M_\p \simeq 0.12~M_\J$.

\subsection{Including Stellar Wind and Radiation Pressure}
\label{sec:method-wind}

The \ext{He} observables of WASP-107b is highly suggestive
of an outflow morphology similar to a comet-like tail, as
the reader will find out shortly
(\S\ref{sec:result-observation},
\ref{sec:result-fiducial-setup}).  We explore two possible
mechanisms that may give rise to the comet-like tail:
stellar wind and radiation pressure. For stellar winds,
although realistic patterns of stellar winds could be very
complicated, we take a very basic approach here. The stellar
wind is injected as a hydrodynamic flow in a simulation,
with two velocity components in the planet frame: (1) a
radial component centered at the host star, and (2) a
headwind due to orbital motion of the planet. The first
component has the radial wind speed as a parameter; the
second one is set to be equivalent to the orbital velocity
of the planet.

The regions of interactions between the stellar wind and the
planet outflow are illustrated by
Figure~\ref{fig:wasp107b-schematic}. Depending on whether
the velocity already becomes supersonic before being
decelerated and deflected by the impinging stellar wind, a
fluid element in the planet outflow should go through the
sonic critical surface twice (part of streamlines in ``Case
A'' and ``Case B'') or once (the complement part of
streamlines in Cases A and B, and all streamlines in ``Case
C''). We note that a streamline in Cases A and B {\it must}
go through the reverse shock when and only when it crosses
the sonic critical surface twice. The first sound crossing
occurs at smaller radii (``inner'' sonic surface) as the
fluid element is accelerated by the thermal pressure
gradient. However, as the upwind part is impinged by the
stellar wind, the streamline travels through a reverse shock
and is decelerated to subsonic. The fluid element then
changes its direction, and become deflected to move in the
night side. The confinement due to stellar wind then becomes
weaker, and the fluid element is allowed to expand like in a
de Laval nozzle before eventually becoming supersonic again
at a second (``outer'') sonic surface. Other streamlines go
through the inner sonic surface only, and never touches the
outer sonic surface or the reverse shock front.

Stellar winds impose more stringent Courant-Friedrichs-Lewy
conditions: each model takes $\sim 51~\hr$ to run on a
40-core, 4-GPU computing node of the Popeye-Simons Computing
Cluster. To accelerate the convergence of our simulations,
in addition to the ``adaptive coarsening'' technique in
Paper I, we also adopt a two-step scheme of simulation: (1)
turn on hydrodynamics, thermochemistry and radiative
transfer, run the simulation for $\sim 15~\tau_\dyn$ until
the model almost reaches the quasi-steady state without
stellar winds; (2) turn on the stellar wind and continue the
simulation for $\gtrsim 200~\tau_\dyn$ until the final
quasi-steady state is reached. The dynamical timescale for a
$T\sim 10^4~\K$ photoevaporative outflow around WASP-107b is
estimated as $\tau_\dyn\sim 4.8\times 10^3~\s$:
\begin{equation}
  \label{eq:t-dyn}
  \tau_\dyn % \sim \dfrac{R_s}{c_s}
  \sim \dfrac{G M_\p}{c_s^3} \sim 1.2\times 10^3~\s\times
  \left(\dfrac{M_p}{10~M_\oplus}\right)
  \left(\dfrac{T}{10^4~\K}\right)^{-3/2}\ .
\end{equation}
To ensure that the simulations are not limited by the
spatial resolution of simulation grid, we ran a model with a
higher resolution
($N_{\log r}\times N_\theta\times N_\phi = 192\times
192\times 128$) and the results are most identical to that
of our standard grid (Figure~\ref{fig:wasp107b-profile}).

Another potentially influential factor is the radiation
pressure. We added this effect to our ray-tracing radiative
transfer procedures by explicitly computing the momenta
deposited by the photons absorbed or scattered. As
\S\ref{sec:discussion-rad-pre} will discuss, the radiative
pressure on \ext{He} transitions plays a very minor role on
the overall dynamics and \ext{He} observables.

\begin{figure*}
  \centering
  \hspace*{-0.1in}
  \includegraphics[width=7.3in, keepaspectratio]
  {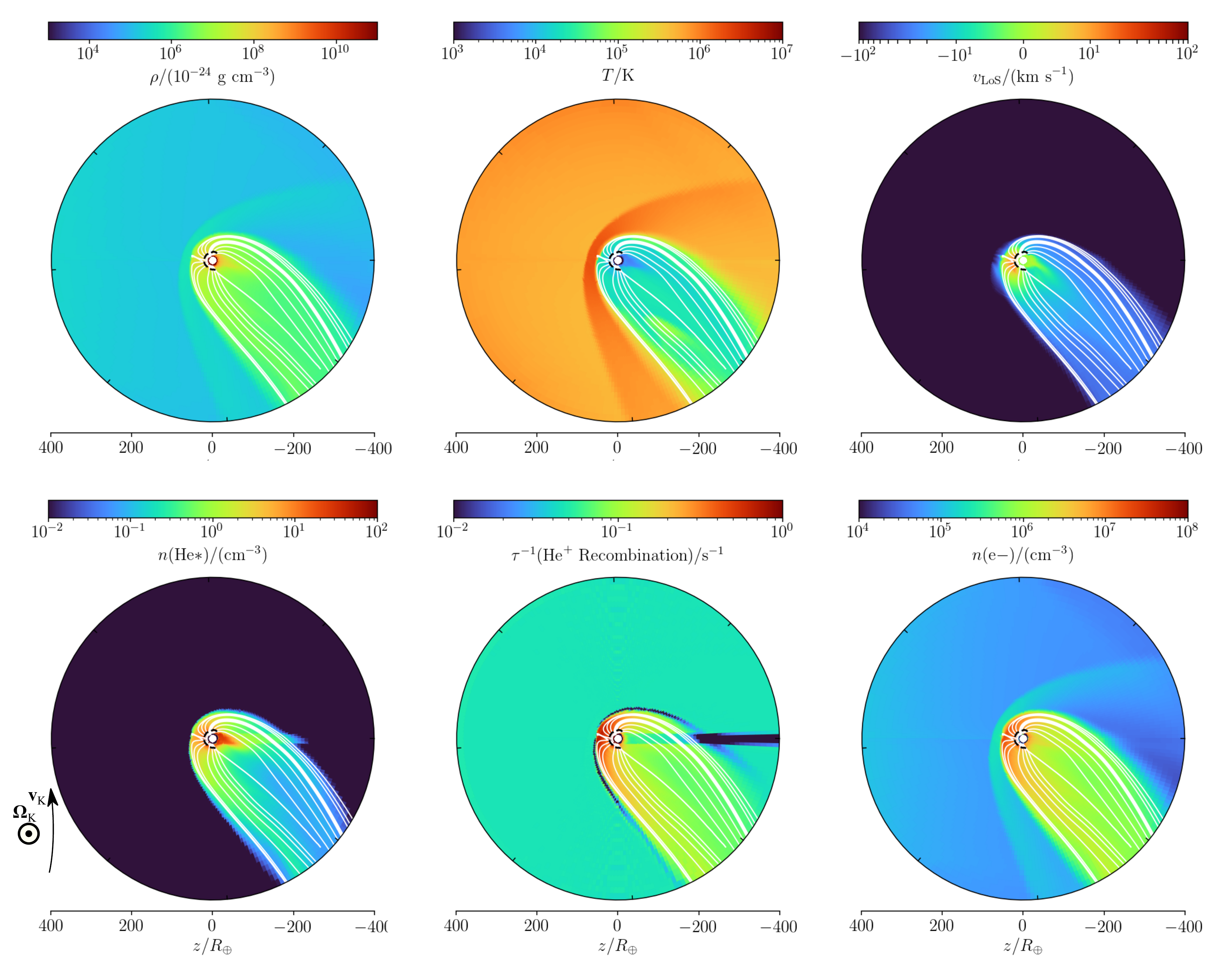}
  \caption{Key hydrodynamics and chemistry profiles of the
    simulation for WASP-107b (fiducial model 107-0) in the
    quasi-steady-state, averaged through the last
    $15~\tau_\dyn$. Stellar radiation comes from the left of
    the plot, and the orbital angular velocity
    $\mathbf{\Omega}_\K$ points out of the paper plane; the
    Keplerian motion of the planet is upwards
    $\mathbf{v}_\K$.
    % Colormaps show the mass density $\rho$
    % ({\bf upper left panel}), temperature $T$ ({\bf upper
    % middle}), line-of-sight velocity $v_\los$ ({\bf
    % upper right}; the value is measured at mid-transit),
    % \ext{He} number density $n(\ext{He})$ ({\bf lower
    % left}), inverse timescale of recombination \ext{He}
    % formation ({\bf lower middle}, defined as formation
    % rate normalized by $n(\ext{He})$), and free electron
    % number density $n(\e^-)$ ({\bf lower right}).
    White streamlines are separated at the {\it wind base}
    by $\Delta \theta = \pi / 16$. Note that only the
    inner sonic surface is shown (in dashed black curve;
    see also the Case A in
    Figure~\ref{fig:wasp107b-schematic}). The heavy
    streamline are the reference lines on which the
    profiles are plotted in
    Figure~\ref{fig:wasp107b-profile}). }
  \label{fig:wasp107b-slice} 
\end{figure*}

\begin{figure}
  \centering
  \includegraphics[width=3.3in, keepaspectratio]
  {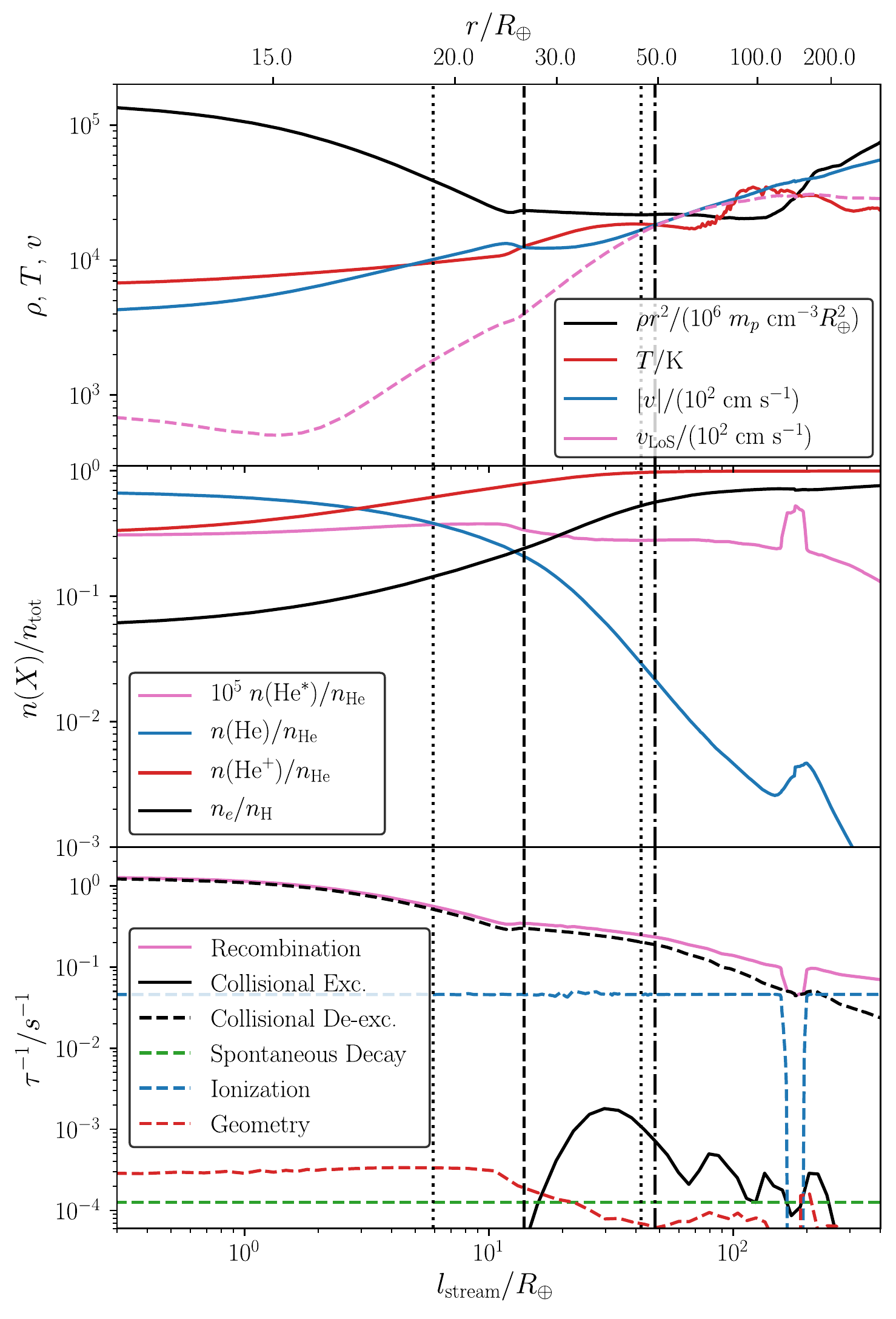}
  \caption{ Key quantities of our fiducial model for
    WASP-107b (Model 107-0), time-averaged over the last
    $15~\tau_\dyn$, along the reference streamlines (plotted
    as the heavy streamline in
    Figure~\ref{fig:wasp107b-slice}). The top abscissa is
    the radial coordinate $r$ corresponding to the curve
    length along the streamline on the bottom abscissa.  As
    explained by \S\ref{sec:method-wind} and the Case A in
    Figure~\ref {fig:wasp107b-schematic}, the fluid elements
    along this streamline experience two smooth crossings of
    sonic speed (indicated by vertical dotted lines; note
    that the right of them is close to the dash-dotted line
    for the Roche radius) and a reverse shock (indicated by
    vertical dashed lines) along the streamline.  }
  \label{fig:wasp107b-profile}
\end{figure}

\begin{figure}
  \centering
  \includegraphics[width=3.4in, keepaspectratio]
  {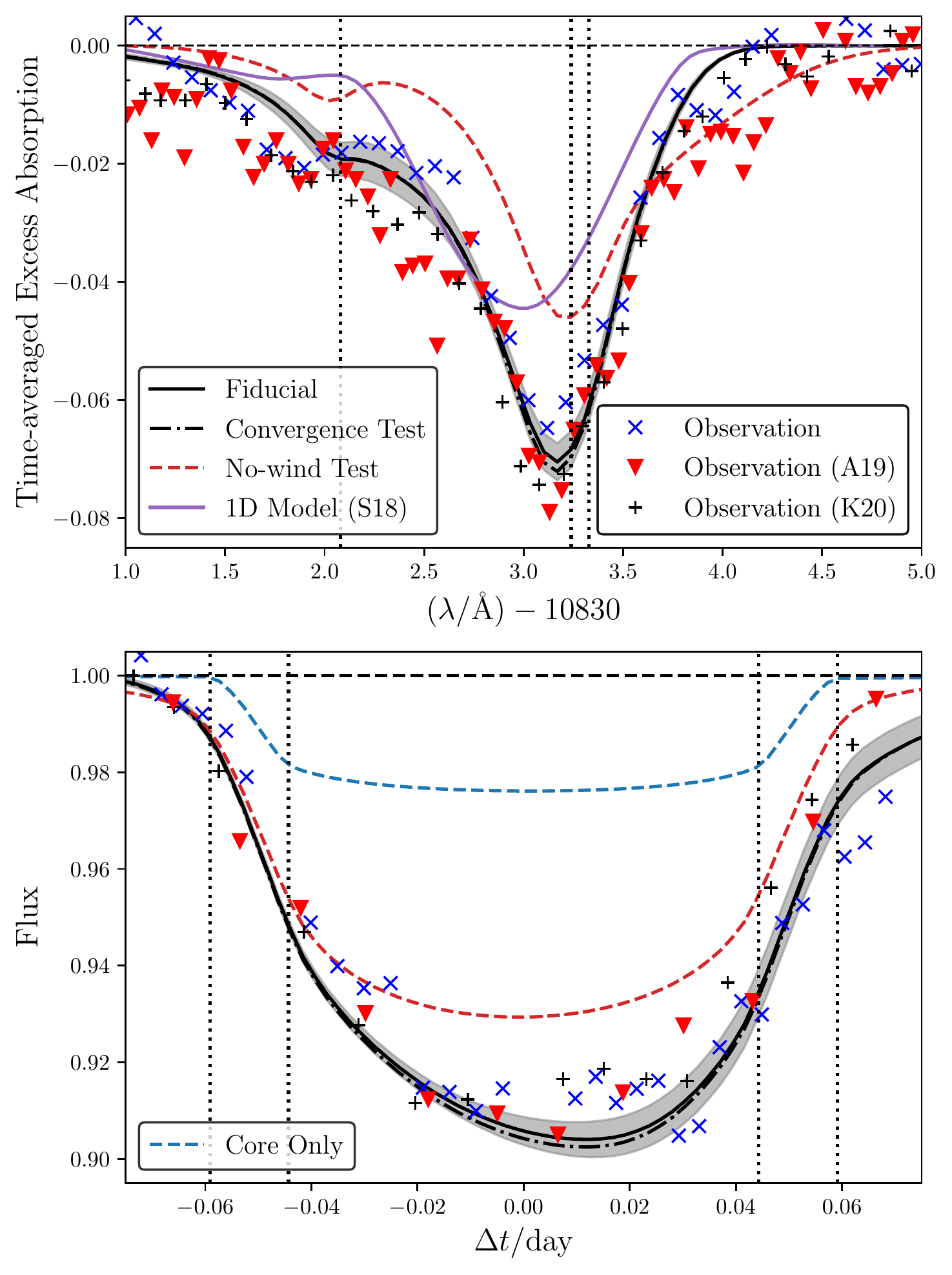}
  \caption{ The observed and synthesized line profiles and
    light curves for the fiducial model of WASP-107b and its
    corresponding test models. Curves for all models
    indicate the averages of the spectral profile and light
    curves over the last $15~\tau_\dyn$ (note that they are
    obtained by calculating the profiles for each snapshot
    {\it before} taking the average, {\it not} by evaluating
    the profiles for the time-averaged model). Grey shades
    around the curves for the fiducial model in both panels
    express the range of two-times standard deviation
    ($2\sigma$) over the last $15~\tau_\dyn$. The
    observation results in \citet{2020AJ....159..115K} are
    denoted by black ``+'' and labeled with ``Observation
    (K20)''; those marked with red triangles and
    ``Observation (A19)'' presents the results in
    \citet{2019A&A...623A..58A}. The 1D spherical symmetric
    model in \citet{2018Natur.557...68S} is included for
    reference, marked by the magenta curve with label ``1D
    Model (S18)'' in the upper panel. For reasonable
    comparisons, the window function for light curves is
    taken to be the same as \citet{2020AJ....159..115K},
    i.e. a top-hat function which has unitary value in
    $(\lambda/\ang)\in [10833.05, 10833.48]$ and zero
    elsewhere (note that the window in
    \citet{2019A&A...623A..58A} is different,
    $(\lambda/\ang)\in [10832.80, 10833.55]$). Vertical
    dotted lines in the upper panel shows the centers of the
    three \ext{He} lines, and mark the start/end of the
    nominal ingress/egress ($t_\roman{i}$ through
    $t_\roman{iv}$) in the lower panel. }
  \label{fig:wasp107b-spec}
\end{figure}

\begin{figure*}
  \centering
  \includegraphics[width=7.0in, keepaspectratio]
  {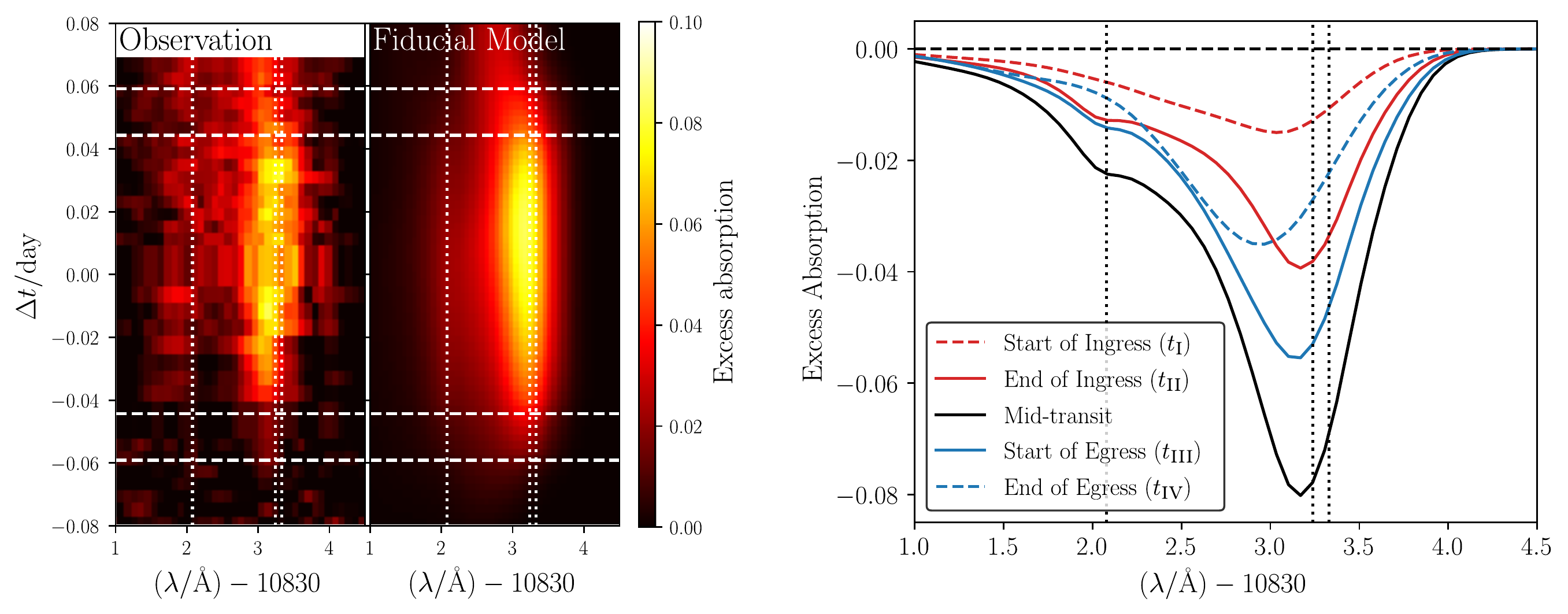}
  \caption{{\bf Left panel}: Comparison of the excess
    absorption features, as functions of transit time
    $\Delta t$ and wavelength $\lambda$, between the
    observation data (left panel) and the synthesized
    results (right panel). The simulation used to produce
    the right panel is the time-averaged fiducial Model
    107-0 (averaging is taken in the same way as
    Figure~\ref{fig:wasp107b-spec}, viz. averaging {\it
      after} the excess absorption map being obtained for
    each snapshot concerned). Both panels already have the
    wavelength shifts by orbital motion subtracted.  Three
    vertical dotted lines in each panel indicate the the
    line centers. Four horizontal dashed lines indicate the
    starts (ends) of the ingress (egress). {\bf Right
      panel}: transmission spectra at different stages of a
    transit, based on the fiducial Model 107-0. }
  \label{fig:wasp107b-heatmap}
\end{figure*} 

\begin{figure*}
  \centering
  \includegraphics[width=7.2in, keepaspectratio]
  {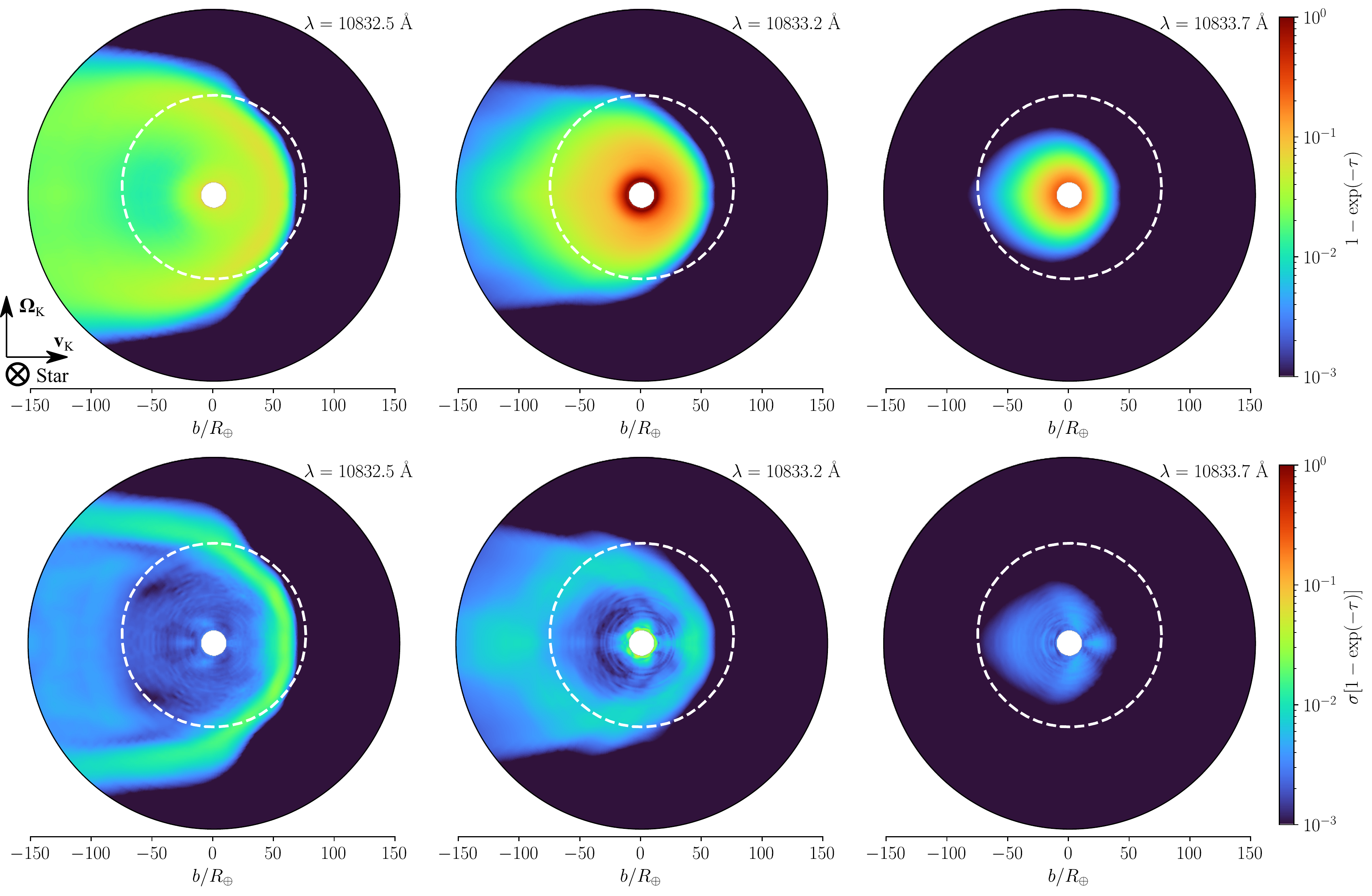}
  \caption{ {\bf First row}: extinction $[1-\exp(-\tau)]$
    for Model 107-0 with time-averaging over the last
    $\sim 15~\tau_\dyn$ (note that the averaging is computed
    {\it after} the extinction function is calculated for
    {\it each} snapshot concerned) at three representative
    wavelengths
    $(\lambda/\ang)\in \{10832.5, 10833.2, 10833.7\}$ near
    the \ext{He} transitions. The profiles are measured at
    mid-transit. {\bf Second row} presents the
    root-mean-squared variations of extinction
    ($\sigma [1-\exp(-\tau)]$). Plotted up to the impact
    parameter $b=150~R_\oplus$. The white dashed circles
    presenting the host star size have
    $R_*\simeq 74.2~R_\oplus$ and orbit impact parameter
    $b\simeq 0.07~R_*$. }
  \label{fig:wasp107b-ext}
\end{figure*}

\begin{deluxetable}{lr}
  \tablecolumns{2} 
  \tabletypesize{\scriptsize}
  \tablewidth{0pt}
  \tablecaption
  {Properties of the fiducial model for WASP-107b
    \label{table:wasp107b-fiducial} } \tablehead{
    \colhead{Item} & \colhead{Value} } \startdata
  Simulation domain & \\
  Radial range& $10.21~R_\oplus \le r \le \ 400~R_\oplus$\\
  Latitudinal range & $0\le\theta\le\pi$ \\
  Azimuthal range & $0\le\phi\le\pi$ \\
  Resolution $(N_{\log r}\times N_{\theta} \times N_{\phi})$ &
  $144\times 128\times 64$ \\   
  \\
  Planet interior$^\dagger$ & \\
  $M_\rcb$ & $38.1~M_\oplus$ \\
  $\mean{r_\eff}$ & $10.6~R_\oplus$ \\
  \\
  Radiation flux [photon~$\cm~^{-2}~\s^{-1}$] & \\
  $2~\eV$ (IR/optical)  & $2.1\times 10^{19}$ \\
  $7~\eV$ (Soft FUV)   & $1.2\times 10^{16}$ \\
  $12~\eV$ (LW)   & $3\times 10^{12}$ \\
  $20~\eV$ (Soft EUV)  & $2\times 10^{13}$ \\
  $40~\eV$ (Hard EUV)  & $6\times 10^{13}$ \\
  $3~\keV$ (X-ray)  & $2\times 10^{12}$ \\
  \\
  Initial abundances [$n_{\chem{X}}/n_{\chem{H}}$] &
  Same as Paper I \\
  Dust/PAH properties &
  Same as Paper I \\
  \\
  Stellar wind (at $a = 0.055~\au$) & \\
  Density &  $1.7\times 10^{-19}~\g~\cm^{-3}$\\
  Temperature &  $10^6~\K$\\
  Radial velocity & $270~\km~\s^{-3}$\\
  Tangential velocity$^*$ & $106~\km~\s^{-3}$\\
  Abundances$^{\ddagger}$
  [$n_{\chem{X}}/n_{\chem{H}}$] & \\
  $\e^-$ & $\simeq 1.1^{**}$ \\
  \pos{H} & 1.0\\
  \pos{He} & 0.1\\
  \pos{O} & $3.2 \times 10^{-4}$\\
  \pos{C} & $1.4 \times 10^{-4}$\\
  \pos{S} & $2.8 \times 10^{-5}$\\
  \pos{Si} & $1.7 \times 10^{-6}$\\
  \enddata
  \tablecomments{ $\dagger$: Mass and transit
    radius of the planet core; see also Appendix A of Paper
    I for details of planet core setups.
    \\
    $\ddagger$: Because the wind injection has a
    star-centered geometry, these values are calibrated at
    the planetary orbit, viz. $a = 0.055~\au$.
    \\
    $*$: Head wind component; same as the Keplerian
    velocity.
    \\
    $\ddagger$: Calibrated at the domain boundary of wind
    injection.
    \\
    $**$: Electrical neutrality is guaranteed.  }
\end{deluxetable}

\section{WASP-107\lowercase{b}: \\an outflow shaped by
  stellar winds}
\label{sec:results}

\subsection{Observation Results}
\label{sec:result-observation}

Quantitatively, we directly compare our synthetic
observations to \ext{He} line profiles and light curves.
The \ext{He} transmission spectrum of WASP-107b is both
spectrally and temporally resolved with different
spectrographs: NIRSPEC and CARMENES
\citep{2018Natur.557...68S, 2019A&A...623A..58A,
  2020AJ....159..115K}. We also carried out an independent
analysis of the transit observed from NIRSPEC on the Keck
telescope on April 6th, 2019 \citep[same
as][]{2020AJ....159..115K}. Our data reduction is carried
out using the procedures described in Zhang, Z. et
al. (submitted). \ext{He} transits have been observed by
several different instruments and different transit events
(Figure~\ref{fig:wasp107b-spec}). Despite the differences in
instrumental characteristics and reduction pipelines,
previous \ext{He} observations agree with each other on the
following features: (1) a non-Keplerian, blueshifted
($\sim 2-3~\km~\s^{-1}$) line profile; (2) a line ratio that
deviates from the 1:3:5 (viz. the quantum degeneracies;
apparently 1:8) among three \ext{He} transitions; (3)
elongated ingress and egress timescales in \ext{He} transit
light curves. Finally, in the temporally resolved
observation \citep{2020AJ....159..115K}, there is a hint of
asymmetry in the outflow morphology that the egress lasts
longer than the ingress while the egress is more blueshifted
compared to the mid-transit (see also
Figure~\ref{fig:wasp107b-heatmap}).

\subsection{Fiducial Model Setup and Necessity of Stellar Winds}
\label{sec:result-fiducial-setup}

As noted in Paper I, consistent 3D hydrodynamic simulations
with all physics are not sufficiently fast for a
comprehensive exploration of the parameter space using
Markov Chain Monte Carlo or even just gradient
descent. Instead, we adopt the parameters based on the
reported stellar and planetary conditions. The optical and
near-infrared fluxes (represented by the $h\nu = 2~\eV$ bin)
simply accord with the host star radius and effective
temperature. The high energy SED of the host star is much
more uncertain. Guided by the results in our Paper I,
i.e. how various high-energy radiation bins affect the
atmospheric outflow and \ext{He} observables, we hand-tuned
the high-energy SED until reasonable agreement with the
observations. The resulting high-energy SED is fairly
typical or slightly more active than a K5 star
\citep{1992A&A...264L..31G, 2016ApJ...820...89F,
  2016ApJ...824..101Y, 2016ApJ...824..102L,
  2017ApJ...843...31Y, 2019ApJ...881..133O}, but we also
note that WASP-107b has a relatively short rotation period
of 17 days, and the $\sim 600~{\rm Myr}$ host star has a
strong chromospheric S index (0.89)
\citep{2017AJ....153..205D}.
   
In addition to tuning the high-energy SED, we found it
crucial to include stellar wind components in our
simulations to adequately reproduce the existing \ext{He}
observations. We first run a series of no-wind model as we
did in Paper I for WASP-69b. The resultant outflow is
largely spherically isotropic, without prominent comet-like
tail. Figure~\ref{fig:wasp107b-spec} presents the \ext{He}
line profile and light curves produced in the no-wind
fiducial model (other parameters are the same as those in
Table~\ref{table:wasp107b-fiducial}). The overall equivalent
width of the line profile is weaker
($\mean{W_\lambda}\simeq (7.6\pm 0.2)~\ang$ in the fiducial
model, but $\mean{W_\lambda}\simeq 5.0~\ang$ in the no-wind
model). More importantly, the line profile does not have a
stronger blueshifted wing, which has been validated by
several different instruments.%  We will explain in more
% detail shortly that in the windy model,
The impinging stellar wind stagnates the outflow flowing
towards the host star (i.e. the redshifted wing), pushing
the outflow to the night-side (the blueshifted wing). The
transit light curve in the no-wind model is also much more
symmetric compared due to the lack of comet-like tail
trailing behind the planet. In short, a no-wind model
struggles, if ever possible, to reproduce the various
\ext{He} observations of WASP-107b. On the other hand, the
windy model successfully reproduced observed \ext{He} line
profiles and light curves. We now examine the windy model in
detail.

\subsection{Fiducial Model Results}
\label{sec:results-fiducial}

Figure~\ref{fig:wasp107b-slice} presents the simulation
results of our fiducial model, with a stellar wind
modulating the outflow morphology. Each panel centers at the
planet's frame; the host star is located to the left of the
this plot, and the stellar wind impinges on the
photoevaporative outflow at an angle given by the ratio
between the radial velocity component and the headwind
(orbital) component. A bow shock of Mach number $\sim 2.0$
forms above the day-side of the planet, heating the
downstream flow to $T\sim 4\times 10^6~\K$ (second panel of
Figure~\ref{fig:wasp107b-slice}). As shown by the
streamlines, the photoevaporative outflow from the planet is
stalled at the day-side and directed by stellar wind to the
night-side, forming a prominent tail trailing the planet's
orbital motion.

Along the thick streamline in
Figure~\ref{fig:wasp107b-slice} (which qualitatively
resembles the ``typical'' streamline in
Figure~\ref{fig:wasp107b-schematic}), we plot the spatial
variation of key hydrodynamic and thermochemical quantities
both as a function of radius from the planet core and the
arc length along the streamline in
Figure~\ref{fig:wasp107b-profile}. Looking at the outflow
velocity first, a fluid element traveling on the streamline
may experience multiple sound crossings, as is shown by the
Case A in Figure~\ref{fig:wasp107b-schematic} (see also
\S\ref{sec:method-wind}). We note that this effect is
strongest at the upwind side where outflow is impinged by
stellar wind; it is much weaker elsewhere and the reverse
shock may no longer exist. The lower panel of
Figure~\ref{fig:wasp107b-profile} indicates that the
population of \ext{He} is determined primarily by the
equilibrium between recombination excitation and collisional
de-excitation at smaller radii. At larger radii
($\gtrsim 10^2~R_\oplus$), photoionization by soft FUV
photons starts to take over. This is very similar to our
results in Paper I for WASP-69b.

The middle panel of Figure~\ref{fig:wasp107b-profile} shows
another interesting feature.  When the streamline crosses
the the planet's shadow, photoionization due to soft FUV
from the host star vanishes. However, recombination
excitation continues in the shadow of the planet and thus
creates a local bump of higher \ext{He} abundances. We call
this effect the ``shadow tail''. Such a feature is difficult
to observe in transits as it always hides in the planet's
shadow. If one is ever able to resolve this photoevaporative
outflow, one may see the shadow tail emitting more strongly
in \ext{He} compared to other parts of the outflow. Such a
``two-tail'' morphology is remarkably similar to the dust
and ion tails of a comets, although with different
underlying physics.

Figure~\ref{fig:wasp107b-ext} offers a transverse view of
the outflow, taken from the perspective of an observer
looking into the host star (outlined as the white
dashed-line) during a transit. The planet is moving towards
the right-hand-side, and is instantaneously close to the
center of the host star. These plots illustrate the spatial
distribution of extinction at three characteristic
wavelengths of the \ext{He} transition. At the bluer
wavelength $\lambda = 10832.5~\ang$, which falls into the
``valley'' between the $J=0$ transition and two blended
$J=1, 2$ transitions, the extinction is dominated by
blueshifted materials in the tail trailing behind the
planet. On the red wing $\lambda = 10833.7~\ang$, the tail
is much less relevant, and the extinction are generated by
materials much closer to the planet. Again, this is
attributed to the kinematics and morphology: the planetary
outflow is impinged by the stellar wind, and the outflow on
the day-side or the headwind direction is stagnated.

Finally, we remind the reader that due to its large
aperture, Keck/NIRSPEC is able to resolve the \ext{He} both
spectrally and temporally. Figure~\ref{fig:wasp107b-heatmap}
presents the observed and simulated variation of \ext{He}
absorption in time and wavelength space together as a
heatmap. The vertical dotted lines are the rest-frame
wavelengths of three \ext{He} transitions, the horizontal
dashed lines are the expected $t_\roman{i}$ to
$t_\roman{iv}$ (starts/ends of the ingress/egress) of the
planet's transit. Our simulation with stellar winds
successfully reproduces the key feature: since most of the
\ext{He} is produced by the comet-like tail trailing the
planet (Figure~\ref{fig:wasp107b-ext}), the strongest
absorption occurs after mid-transit, while the pre-ingress
absorption is much weaker than the post-egress one. This
feature is also manifested by the asymmetry in the transit
light curve (Figure~\ref{fig:wasp107b-spec}). The outflow
morphology loses spherical symmetry; the part ahead of the
planet and behind the planet are both blueshifted. As a
result, the \ext{He} line profiles seen pre-ingress and
post-egress are both more blueshifted compared to the
mid-transit line profile. This is again consistent with the
NIRSPEC observation.

\begin{figure*}
  \centering
  \hspace*{-0.1in}      
  \includegraphics[width=7.3in, keepaspectratio]
  {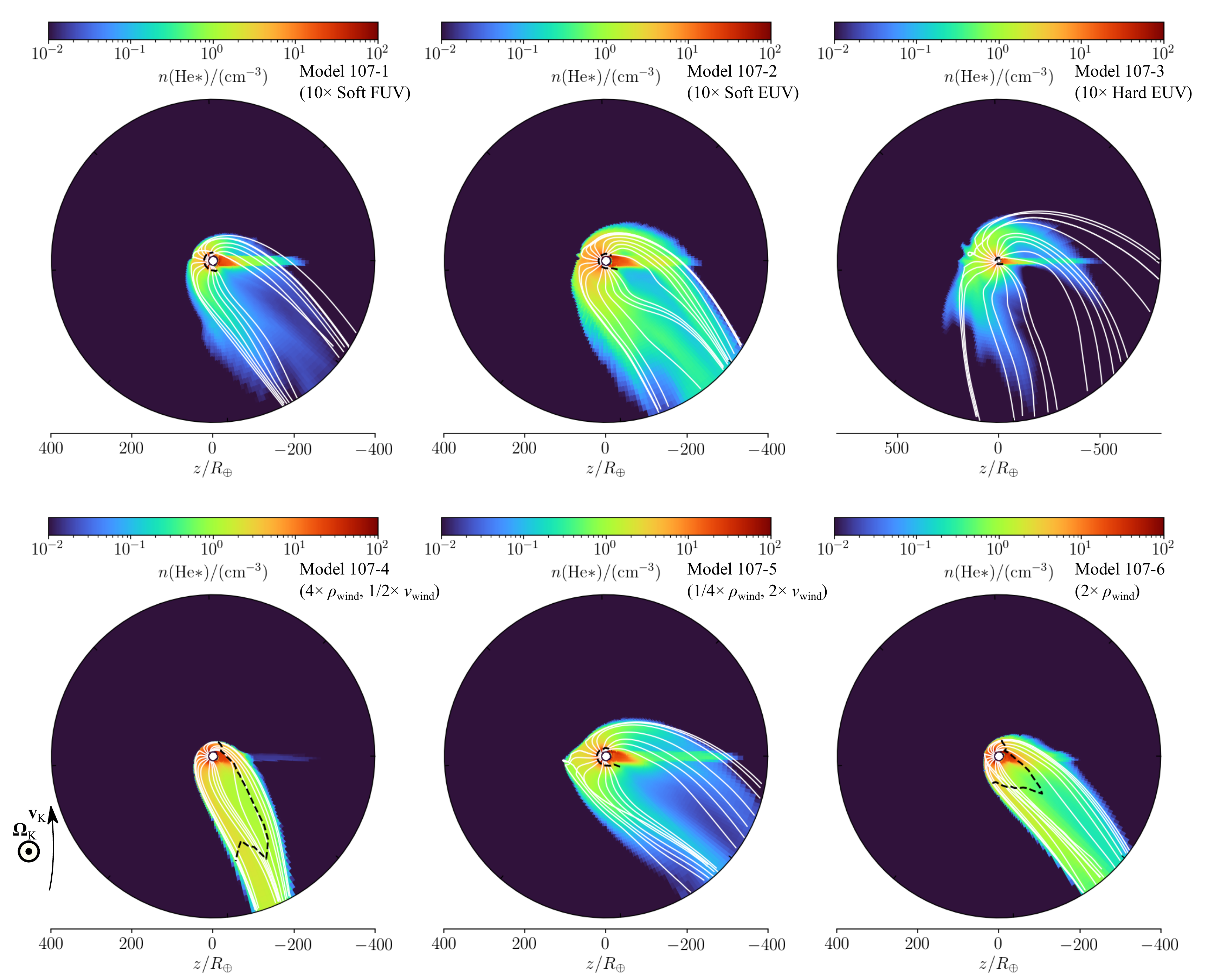}
  \caption{\ext{He} number density profiles in the orbital
    plane, with streamlines (spaced with the same scheme as
    Figure~\ref{fig:wasp107b-slice}) and sonic surfaces
    overlaid. The panel for Model 107-3 has different
    spatial scales. Some lines appear to overlap with other
    lines or to terminate before reaching the outer boundary
    because the streamlines are projected. Note that Models
    107-4 and 107-6 comply with the Case C in
    Figure~\ref{fig:wasp107b-schematic}. }
  \label{fig:wasp107b-var-slice} 
\end{figure*}
 
\begin{figure*}
  \centering
  \includegraphics[width=7.1in, keepaspectratio]
  {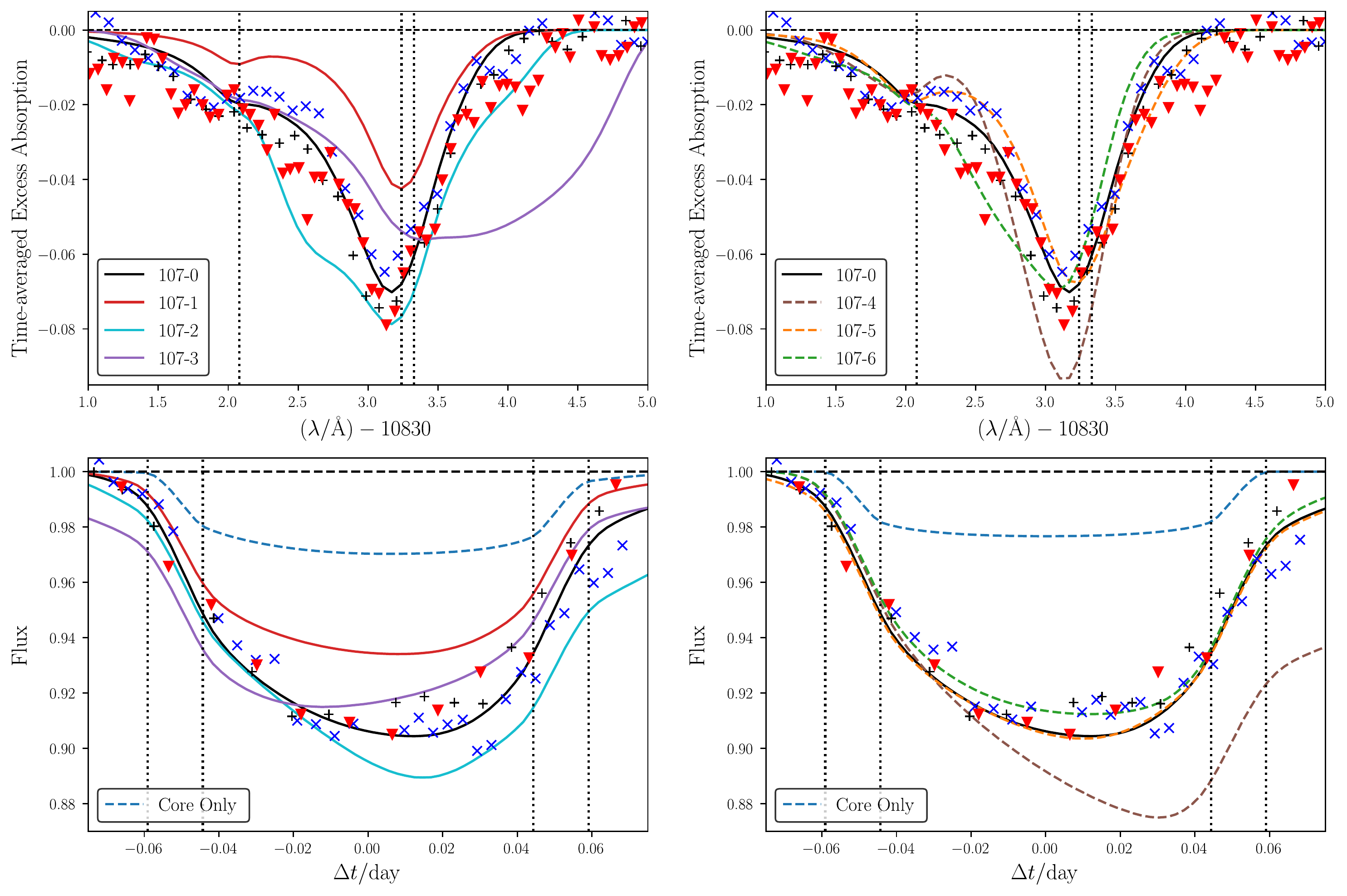}
  \caption{Similar to Figures~\ref{fig:wasp107b-spec} but
    for the models in \S\ref{sec:parameter-wind} and
    Table~\ref{table:wasp107b-var}. {\bf Left column}
    compares Models 107-1 through 107-3 to the fiducial
    107-0, and the {\bf right column} does the comparison
    for Models 107-4 through 107-6 to 107-0. The window
    function for the light curves are the same as the one
    used to yield Figure~\ref{fig:wasp107b-spec}.  }
  \label{fig:wasp107b-var-spec}
\end{figure*}
 
\renewcommand{\arraystretch}{1.2}
\begin{deluxetable*}{lccccc}
  \tablecolumns{6} 
  \tabletypesize{\scriptsize}
  \tablewidth{0pt}
  \tablecaption{Results of various models, based on 
    the fiducial model for WASP-107b 
    \label{table:wasp107b-var} }
  \tablehead{
    \colhead{Model} \vspace{-0.25cm} &
    \colhead{Description} &
    \colhead{$\dot{M}$} &
    \colhead{$\mean{W_\lambda}$} &
    \colhead{$\Delta v_{\rm peak}$} &
    \colhead{FWHM}
    \\
    \colhead{ } &
    \colhead{ } &
    \colhead{($10^{-9}~M_\oplus~\yr^{-1}$)} &
    \colhead{($10^{-2}~\ang$)} &
    \colhead{($\km~\s^{-1}$)} &
    \colhead{($\km~\s^{-1}$)} 
    }
  \startdata
  107-0 & Fiducial & $1.02\pm 0.01$ & $7.6\pm 0.2$
  & $-3.5$ & $22.7$ \\
  \hline
  107-1 & $10\times $ Flux at $h\nu = 7~\eV$ &
  $1.09\pm 0.01$ & $3.7\pm 0.1$ & $-1.7$ & $17.2$ \\
  107-2 & $10\times $ Flux at $h\nu = 20~\eV$ &
  $1.23\pm 0.01$ & $11.3\pm 0.9$ & $-3.7$ & $34.6$ \\
  107-3 & $10\times $ Flux at $h\nu = 40~\eV$ &
  $3.47\pm 0.03$ & $12.1\pm 1.1$ & $~\,$ $2.4$ & $55.1$ \\
  \hline
  107-4 & $4\times \rho_\wind$, $1/2\times v_\wind$ &
  $0.98\pm 0.02$ & $8.6\pm 0.1$ & $-4.5$ & $20.4$ \\
  107-5 & $1/4\times \rho_\wind$, $2\times v_\wind$ &
  $1.02\pm 0.01$ & $7.4\pm 0.4$ & $-2.3$ & $22.6$ \\
  107-6 & $2\times \rho_\wind$ &
  $0.99\pm 0.02$ & $8.2\pm 0.1$ & $-5.6$ & $28.2$ \\
  \enddata
  \tablecomments{The values and errors are time
    averages and three times the standard deviations
    ($3\sigma$), respectively; the time averages are taken
    over the last $25~\tau_\dyn$ of the simulations. }
\end{deluxetable*}

\section{Parametric Studies}
\label{sec:var-params}

The previous section elaborates the fiducial model for
WASP-107b. We found that, by including stellar winds, the
simulation quantitatively reproduces various observed
features of the \ext{He} transits. This section investigates
how the outflow morphology and \ext{He} observables are
affected by each model parameter
(Figures~\ref{fig:wasp107b-var-slice} and
\ref{fig:wasp107b-var-spec}). We set a series of models,
each differs from the fiducial model by one parameter only
unless specifically indicated. Table~\ref{table:wasp107b-var}
and Figure~\ref{fig:wasp107b-var-spec} summarize the models
and their corresponding key \ext{He} observables.

\subsection{High-Energy Radiation Fluxes}
\label{sec:var-flux}

Paper I found that \ext{He} transit observation are
primarily determined by soft FUV (which destroys the
\ext{He} state) and the EUV bands (which is most important
in driving the photoevaporative outflow; interested readers
are referred to Paper I for more details); the Lyman-Werner
and X-ray bands play secondary roles. We found very similar
results for WASP-107b after including stellar winds, by
running a separate test simulation excluding X-ray photons
(not presented in this paper) that yields almost identical
morphology and observables to the fiducial one. In Model
107-1 through 107-3, we raise stellar fluxes in the soft FUV
($h\nu = 7~\eV$), soft EUV ($20~\eV$), and hard EUV
($40~\eV$) bands, respectively.

Model 107-1 increases the soft FUV flux by a factor of
10. As we have noted earlier, soft FUV does not inject
significant heat into the atmosphere and has minor effects
in driving the planetary outflow. Its mass loss rate is very
similar to that in the fiducial model
[$(\dot{M}/10^{-9}~M_\oplus~\yr^{-1}) = 1.08$ versus
$1.02$]. Soft FUV is nonetheless capable of photoionizing
the \ext{He} state efficiently, slicing the equivalent
widths by a factor of 2
[$(\mean{W_\lambda}/10^{-2}~\ang = 3.7$ versus 7.6). Models
107-2 and 107-3 raise the soft and hard EUV flux level by a
factor of 10, respectively. As expected, intense EUV fluxes
boosted the photoevaporation rates, indicated by the
stronger mass loss rate and equivalent width. What is
curious, however, is that in 107-3, the planetary outflow is
so strong that it is push back the bow shock with stellar
wind on the day-side to much higher altitudes. In order to
fully contain the pushed-back contact surface and the bow
shock, we use $r_{\rm out} = 800~R_\oplus$ for the outer
radial boundary of 107-3. \footnote{We carried out another
  simulation run that has $r_{\rm out} = 400~R_\oplus$ (same
  as other runs) for the radial boundary, while all other
  physical conditions remain the same as Model 107-3. It is
  confirmed that this test run (not shown in the paper) has
  all quantitative characteristics almost identical to
  107-3, despite that its mass loss rate is $\sim 5~\%$
  greater.} This effect is so strong that we begin to see
the day-side (pre-ingress) material in redshift, and the
overall \ext{He} line profile is now biased towards the red
wing (Figure~\ref{fig:wasp107b-var-spec}, upper panel). This
strong day-side outflow also manifests as a reversed
asymmetry in the \ext{He} light curve, where the ingress is
more extended and stronger than the egress
(Figure~\ref{fig:wasp107b-var-spec}, lower panel). This
qualitative change in behavior will possibly constrain the
balance of high-energy radiation strengths between the
stellar wind and the planetary outflow.

\subsection{How Stellar Winds Shape Outflows}
\label{sec:parameter-wind}

In hydrodynamic simulations (no magnetic effects included),
the interactions between stellar winds and planetary
outflows should be determined by the wind ram pressure
($\sim \rho_\wind v_\wind^2$), as a function of the density
($\rho_\wind$) and velocity ($\vec{v}_\wind$). Fortunately,
this degeneracy between wind density and velocity can be
potentially broken by orbital motion, i.e. the headwind
component of the stellar wind.

On one hand, Model 107-4 quadruples the density $\rho_\wind$
and halves the velocity $v_\wind$, thus keeping the radial
(from host to planet) component of ram pressure roughly the
same as in the fiducial model. The headwind component,
caused by the Keplerian motion $v_\K$ of the planet, is held
constant here. Lower velocity shear yields a less turbulent
trailing tail with weaker shear instabilities. As a result,
the tail lags behind the planet in both velocity and spatial
sense, producing a more blueshifted line profile
[$(\Delta v_{\rm peak}/\km~\s^{-1})=-4.5$, versus $-3.5$ for
the fiducial], and a more asymmetric light curve. On the
other hand, in Model 107-5 which doubles the velocity but
quartered the density, the tail is now influenced by the
radial component more heavily.  It is primarily directed
towards the night-side, and displays considerably more
vigorous shear instabilities due to greater velocity
shears. The \ext{He} light curve has much weaker post-egress
absorption, while the equivalent width sees more variability
[$(\mean{W_\lambda}/10^{-3}\ang) = 7.4\pm 0.4$, versus
$7.6\pm 0.2$ in the fiducial model]. Finally, Model 107-6,
only increases the density $\rho_\wind$ by a factor of
two. Having the same ratio of $v_\wind/v_\K$ as in the
fiducial model, Model 107-6 produces a tail that has similar
direction as the fiducial model. Increased ram pressure
still pushes the outflow to greater blueshift velocities
[$(\Delta v_{\rm peak}/\km~\s^{-1})=-5.6$ versus $-3.5$].

\section{Discussions}
\label{sec:discussion}

\begin{figure}
  \centering
  \includegraphics[width=3.4in, keepaspectratio]
  {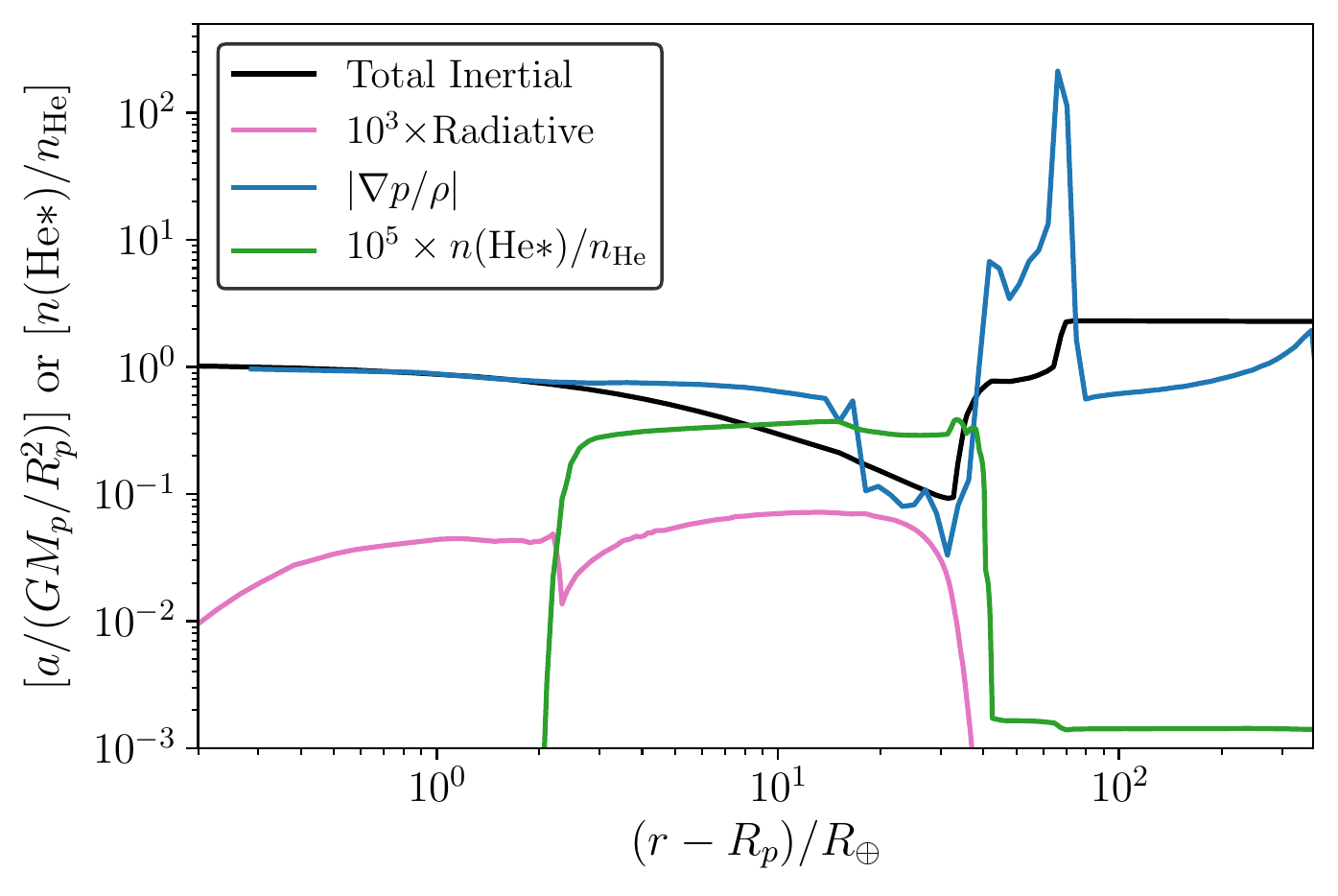}
  \caption{Profiles of acceleration magnitudes (normalized
    by $G M_\p/R_\p^2$) that contains extra photon energy
    bins at \lya and $10833~\ang$ radiated by the host
    star. Inertial acceleration (including planet and
    stellar gravity, centrifugal force, and Coriolis force),
    radiation pressure acceleration (multiplied by $10^3$)
    and pressure gradient acceleration profiles are measured
    along the $\theta = \pi/4$, $\phi = 0$ radial line.  The
    Radiation pressure profile is multiplied by $10^3$ for
    clearer presentation. As a reference, the relative
    abundance of \ext{He} (multiplied by $10^5$) is also
    shown. }
  \label{fig:wasp107b-acc}
\end{figure}

\begin{figure}
  \centering
  \includegraphics[width=3.4in, keepaspectratio]
  {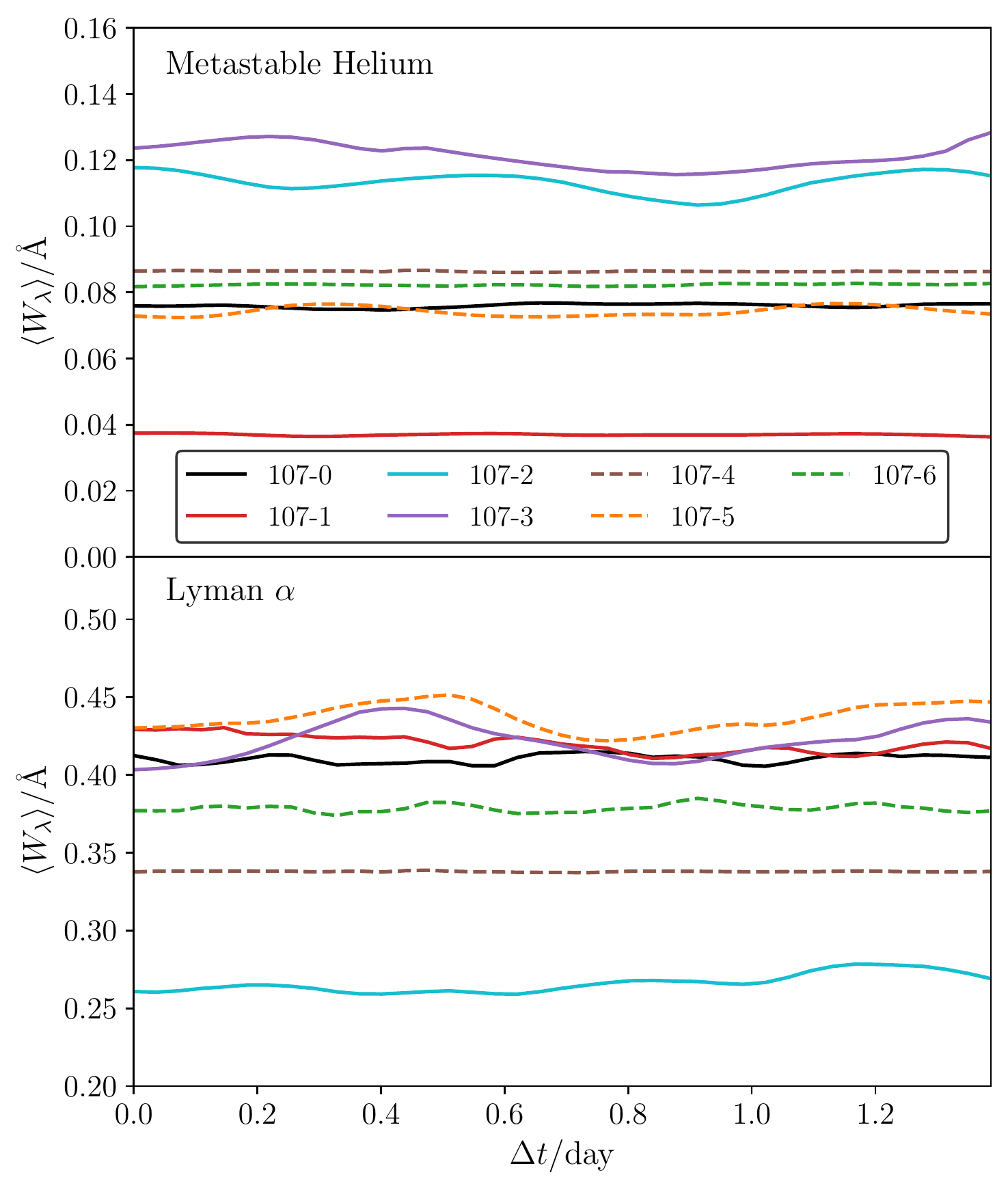}
  \caption{Temporal variation of the equivalent widths of
    the metastable helium absorption lines (upper panel) and
    \lya (lower panel), for the models
    described in Table~\ref{table:wasp107b-var}. Such
    fluctuations are mainly the results of the instabilities
    emerging in the interactions between stellar wind and
    planetary outflows. For each simulation, the last
    $1.4~{\rm day}\sim 25~\tau_\dyn$ of its simulated time
    duration is shown.
  }
  \label{fig:wasp107b-var-ew}
\end{figure}

\begin{figure*}
  \centering
  \includegraphics[width=7.2in, keepaspectratio]
  {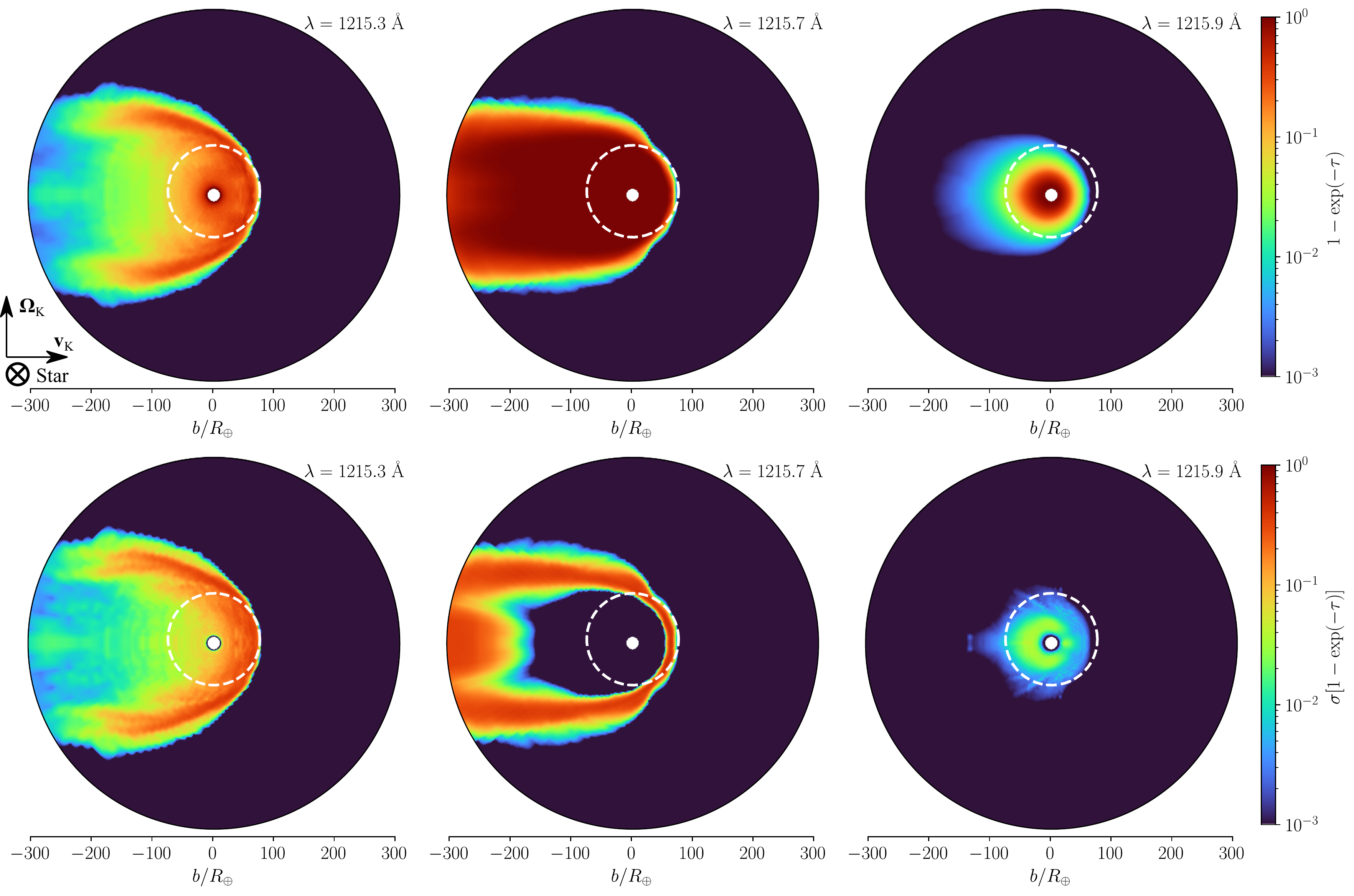}
  \caption{Similar to Figure~\ref{fig:wasp107b-ext} but for
    the wavelengths
    $(\lambda/\ang)\in\{1215.3, 1215.7, 1215.9\}$ near the
    \lya line.}
  \label{fig:wasp107b-lya-ext}
\end{figure*}

\begin{figure}
  \centering
  \includegraphics[width=3.4in, keepaspectratio]
  {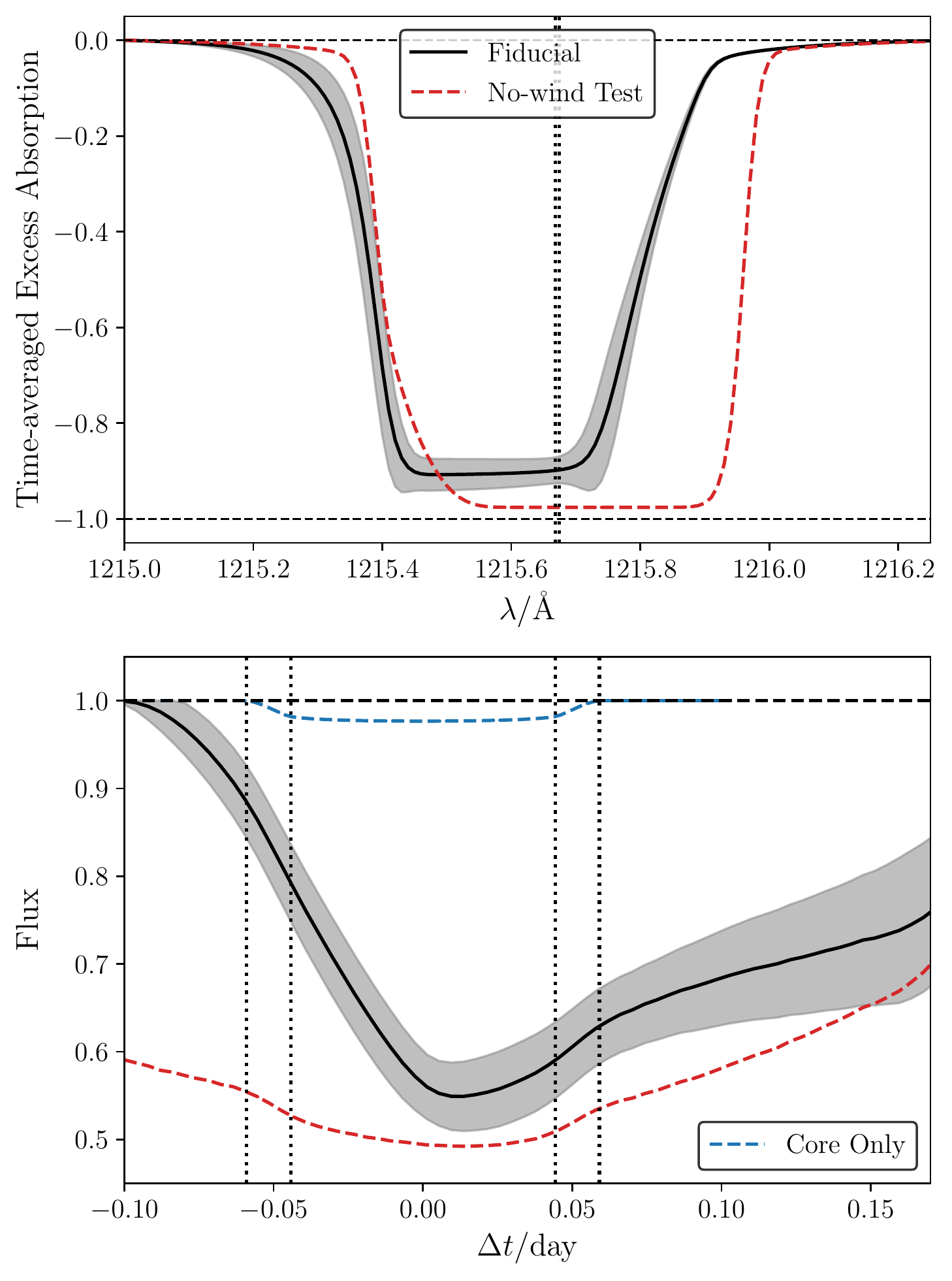}
  \caption{Similar to Figure~\ref{fig:wasp107b-spec} but for
    the \lya synthetic spectra based on Model 107-0 and its
    corresponding test models only. The wavelength window
    for the light curves is
    $(\lambda/\ang)\in [1215.0,1216.25]$. Note that the line
    profiles in the upper panel has a saturated shape but
    the excess absorption does not reach
    $\mean{\Delta\epsilon} = -1$, because (1) the extinction
    by planet core is subtracted from the total absorption,
    and (2) the full ingress of the \lya absorption region
    is delayed by the stellar wind. }
  \label{fig:wasp107b-lya-spec}
\end{figure}

\subsection{Tail: Stellar Wind or Radiation
  Pressure?}
\label{sec:discussion-rad-pre}

As we have shown in the previous sections, the existing
\ext{He} observations of WASP-107b are suggestive of a
comet-like tails that trail the photoevaporating
planet. What is the mechanism that give rise to these tails?
Is it stellar wind or radiation pressure or a combination of
these two effects?  When studying the \lya outflow for
exoplanets, previous studies often attributed transit
asymmetry to radiation pressure \citep[e.g.][]
{2015A&A...582A..65B, 2016A&A...591A.121B}. Recent works in
\ext{He} followed suit: in an EVE simulation of WASP-107b (a
collisionless particle-based simulations;
\citealt{2018Sci...362.1384A, 2019A&A...623A..58A}), the
authors found that their prescription of radiation pressure
accelerates the outflow way too quickly, producing a tail
that is directly pointing away from the night-side of the
planet. Their prescribed effect of radiation pressure is so
strong that the authors had to artificially decrease the
stellar spectrum near the \ext{He} lines by a factor 50 to
achieve reasonable agreement with the observations of
WASP-107b. We argue, in this work, that their particle-based
simulation is not adequate to account for the effect of
radiation pressure on \ext{He} transitions.

We must first clearly distinguish the \lya transition and
the \ext{He} transitions. The lower state of \lya is the
ground state of neutral hydrogen, while \ext{He} (i.e. the
$2^3$S state of helium) is relatively much more difficult to
populate, and has a much lower abundances throughout the
domain of interest ($\lesssim 10^{-5}$,
Figure~\ref{fig:wasp107b-profile}; see also Paper I;
\citealt{2019ApJ...881..133O}). Meanwhile, the product of
Einstein $A$ coefficient and the overall element abundance
for helium is also $\sim 10^2$ times smaller than
\lya. Consequently, the altitude where most \ext{He}
absorption occurs is often much lower than \lya absorption
region. The gas density in these regions is significantly
higher, and the mean free path (MFP) of momentum transfer
are much shorter. In our simulations, we typically see
$n(\pos{H})\gtrsim 10^6~\cm^{-3}$, and an MFP of
$\lambda = 1/[n(\pos{H})\sigma] \lesssim 10^3~\cm$
\citep[see also][]{Thomas_1972, Draine_book}. This is
$\sim 6$ orders of magnitude smaller than the hydrodynamic
length scales. Such a contrast in length scales demands a
hydrodynamic treatment, instead of a collisionless
particle-based one. What is more, the momenta deposited into
\ext{He} by absorption are quickly diluted and thermalized
by more collisions. In contrast, the collisionless
particle-based treatment such as EVE is more appropriate for
simulating the \lya outflow, since the density where most
\lya absorption occurs is much lower: in our simulations, we
have $n(\pos{H})\lesssim 10^{-1}~\cm^{-3}$ and the MFP is
$\lambda\sim 10^{10}~\cm \sim 10~R_\oplus$. Neutral hydrogen
atom there are essentially decoupled from other components
of the gas, and are indeed susceptible to the momentum
deposited by the absorption of \lya. Nonetheless, as far as
the \ext{He} observations are concerned, the radiation
pressure on \lya can be safely ignored.

Figure~\ref{fig:wasp107b-acc} compares the magnitudes of
acceleration (force per unit mass) by radiation pressure
(where most \ext{He} absorption occurs), gas pressure
gradient, and inertial force (including stellar and
planetary gravitation, centrifugal force, and Coriolis
force). Radiation pressure in this model is more than three
orders of magnitude weaker than the pure hydrodynamic
effects. Again, the high density of \ext{He} absorbing
region maintains sufficiently strong momentum coupling of
different species and guarantees quick dilution of the
momenta injected by \lya photons. The photons here heat
rather than push the planetary outflow.

In addition, we set up a verification test simulation to
numerically examine the effect of radiation pressure. We
added two more photon energy bins in our ray-tracing
radiative transfer module: one has $h\nu = 10.2~\eV$ for
\lya, and $h\nu = 1.15~\eV$ for the metastable helium lines
whose flux is of a typical K-type stars
\citep{2005ApJS..159..118W}. For simplicity, we assume that
the cross-sections for photon momentum absorption for these
two energy bins are equal to the line-center value, and we
ignored the dispersion of photons in the frequency space. We
note that these assumptions should only amplify the effects
of radiation pressure. This simulation produced almost
identical result as a no-radiation-pressure model in terms
of hydrodynamic and thermochemical profiles, as well as
\ext{He} observables.

In summary, at least for \ext{He} lines, radiation pressure
only plays a minor role in shaping the outflow morphology
and kinematics. Stellar wind, as we have shown in previous
section, is essential in reproducing various observed
features \ext{He} transitions. We propose that, should
future observation of \ext{He} show strong deviations from a
``quiescent'' line ratio ($\sim 1:8$) or an asymmetry in
line profile or light curve, it may be regarded as an
indication of stellar winds, whose properties can be
constrained with detailed 3D hydrodynamic simulations.

\subsection{Shear Instability}
\label{sec:discussion-si}

At the contact surface that separates the stellar wind and
the photoevaporative outflow, there is region of large
shears in velocity. Shear instabilities are generated here,
manifesting themselves as billowing outflows. We found that
since the sonic surface is located at much lower altitude
(Figure~\ref{fig:wasp107b-schematic}; see also
Figure~\ref{fig:wasp107b-profile}), the outflow mass-loss
rate is not affected by these shear instabilities. If the
host star high-energy SED is held constant, the mass loss
rate stays rather stable ($< 1~\%$ fluctuation) at about
$\dot{M} \simeq 1.02\times 10^{-9}~M_\oplus~\yr^{-1}$ in our
fiducial model of WASP-107b. However, the \ext{He}
absorption takes place at higher altitudes
(Figure~\ref{fig:wasp107b-ext}), and is directly influenced
by shear instabilities. Fluctuations of transit depths have
relative amplitudes $\sim 5~\%$ to $\sim 20~\%$ at different
wavelength, while the typical timescale is
$\sim 0.2~{\rm day}$ (Figure~\ref{fig:wasp107b-var-ew}). In
contrast, in our WASP-69b model in Paper I, shear
instabilities are not excited since stellar wind was not
included. \ext{He} absorption shows a $\lesssim 0.5~\%$
temporal variations.

An interesting prediction associated with shear instability
is that there should be much stronger fluctuations in the
blueshifted wings of \ext{He} line profiles. Moreover, in
the \ext{He} light curve, the variability should be much
stronger after the mid-transit point. The reason is that
shear instability takes time and space to grow: its spatial
extent is still small in the redshifted head part of the
outflow, but fully developed when it enters the blueshifted
tail. In Figure~\ref{fig:wasp107b-spec}, the gray shaded
region indicates the amplitudes of variability due to shear
billows, and clearly demonstrates this fluctuation-induced
asymmetry. Up to the composition of this paper, two \ext{He}
transits have been reported; more transits, preferably from
the same instruments, are needed to evaluate the fluctuation
asymmetries.

\subsection{Line Ratio Probes Kinematics Instead of Density}
\label{sec:discussion-line-ratio}

It has been suggested that the line ratios between three
transition of the \ext{He} can tell us about the density of
the underlying outflow. If all three lines are not
saturated, the line ratios between the triplet should be
proportional to 1:3:5, or their quantum
degeneracies. Considering that the two longer-wavelength
transitions are often blended together thermally and
kinematically, the line ratio should be 1:8. Now if the
density of the outflow is high enough that the transitions
start to saturate, the line ratio may begin to deviate from
1:8 and hint us about the number density of \ext{He} in the
outflow. However, as we have seen in Paper I for WASP-69b,
most of the \ext{He} extinction happens at high-altitude,
low-density regions of $10^{1-2}~R_\oplus$ (see
Figure~\ref{fig:wasp107b-ext}, showing the mid-transit
extinction at three characteristic wavelengths near the
\ext{He} transitions). The lines are far from saturation in
these regions. In other words, the line ratios can {\it not}
directly probe the density distribution at least for
WASP-107b.

We do observe a line ratio of about 1:4 in our fiducial
model of WASP-107b, which is significantly different from
the 1:8 expected by quantum degeneracies. It is noted that
this deviation is primarily caused by the kinematics of the
outflow: the peak due to two longer-wavelength transitions
are blueshifted by the wind by up to
$\sim 20-30~\km~\s^{-1}$, that its blueshifted wing start to
invade the shorter-wavelength transition. We will see in the
next subsection that the line ratio and the exact shape of
the \ext{He} will serve as a probe of the outflow kinematics
and in turn the properties of stellar wind.

Here we present an example that one can semi-quantitatively
constrain the stellar wind density and velocity using the
\ext{He} observation. We shall focus on the contact
discontinuity, i.e. the boundary separating the stellar wind
and the planetary outflow
(Figures~\ref{fig:wasp107b-schematic},
\ref{fig:wasp107b-slice}). The distance of contact
discontinuity from the planet can be estimated by equating
the total pressure of a photoevaporative outflow to the ram
pressure of the stellar wind $\rho_\wind v_\wind^2$:
\begin{equation}
  \label{eq:discusssion-wind-cd}
  \begin{split}
    r_{\rm cd} \sim & \left[ \dfrac{\dot{M}c_s}
      {\pi \rho_\wind v_\wind^2} \right]^{1/2}
    \\
    \sim & ~ 500~R_\oplus\times
    \left( \dfrac{\dot{M}}{10^{-9}~M_\oplus~\yr^{-1}}
    \right)^{1/2}
    \left( \dfrac{T}{10^4~\K}\right)^{1/4}
    \\
    & \times\left(\dfrac{\rho_\wind}
      {10^{-20}~\g~\cm^{-3}}\right)^{-1/2}
    \left( \dfrac{v_\wind}{100~\km~\s^{-1}}\right)^{-1}\ ,
  \end{split}
\end{equation}
where the subscript ``cd'' stands for the contact
discontinuity. In our fiducial model, this estimation yields
$r_{\rm cd}\sim 45~R_\oplus$ with the parameters in
Table~\ref{table:wasp107b-fiducial}. This is reasonably
accurate with visual inspection of
Figure~\ref{fig:wasp107b-slice}. We note that if the stellar
wind intensity is similar to the Sun at an $a = 0.055~\au$
orbit, which roughly has
$\rho_\wind \sim 3\times 10^3~m_p~\cm^{-3}$ and
$v_\wind \sim 300~\kms$ at $\sim 0.05~\au$ (see
\citealt{2018A&A...611A..36V} and references therein),
$r_{\rm cd}\sim 240~R_\oplus$ would be inconsistent with the
observations on WASP-107b. Such stronger stellar wind
experienced by WASP-107b is perhaps not surprising, given
its closer-in orbit, and the relative youth and higher
activity of the host star.

\subsection{Synergy between \lya and \ext{He} Observations}
\label{sec:discussion-lya}

This section briefly discusses how observing transits of an
exoplanets in both \lya and \ext{He} can be synergistic in
helping us understand its atmospheric outflow. First of all,
we reiterate the point made in
\S\ref{sec:discussion-rad-pre} that most \lya absorption
happens at much higher altitudes than \ext{He}
absorption. The low density at higher altitude means that
\lya radiation pressure may start to reshape the morphology
of neutral hydrogen; while the at lower altitude of \ext{He}
absorption, radiation pressure heats rather than pushes the
planetary outflow. This potentially allows us to disentangle
the influence of radiation pressure and stellar wind, and
disentangle the inner and outer parts of the planet outflow.

We produced synthetic line profiles and light curves in the
\lya band for WASP-107b in
Figure~\ref{fig:wasp107b-lya-spec} and
Figure~\ref{fig:wasp107b-lya-ext}. Note that we did not
account for the extinction due to the interstellar
medium. We also note that \lya observations are currently
unavailable for WASP-107b and will probably remain the case
given the distant host and the expected strong UV
extinction. However, we still notice that the line profile
is blueshifted significantly while \lya light curve shows
much more pronounced asymmetry with elongated egress. Both
of these observations are qualitatively similar to other
exoplanets observed in \lya (e.g., GJ 436b; see also
\citealt{2015Natur.522..459E, 2015A&A...582A..65B,
  2016A&A...591A.121B}). It is also curious that radiation
pressure alone could not produce as strong as an asymmetry
produced by stellar wind. The no-wind model produces deep
\lya transit light curve that ingress and egress elongated
to similar extent. Another prediction we can predict from
joint \lya and \ext{He} is that the absorption fluctuation
due to shear instabilities should be stronger in \lya than
in \ext{He}; and \lya fluctuations should lag behind
\ext{He} fluctuations. The reason is, again, most \lya
absorption happen at higher altitude than \ext{He}
absorption. Instabilities grows as it propagates from lower
altitude region to higher altitudes. The shaded bands in
Figure~\ref{fig:wasp107b-lya-spec} illustrate the simulated
variations in \lya absorption.

Some of the points raised here were also suggested by
\citet{2019ApJ...873...89M} who focused hot Jupiters. After
all, transit observations in \ext{He} and \lya, particularly
multiple simultaneous transit observations, will be
instrumental in helping us understand the morphology and
kinematics of atmospheric loss on different spatial
scales. With such observations, we may also acquire more
knowledge about the characteristics and variabilities of
stellar wind from the host star. Hybrid models combining
hydrodynamics with particles-based simulations are perhaps
mandatory to fully characterize the various species of the
outflow at highest altitude especially neutral hydrogen.

\section{Summary}
\label{sec:summary}

This paper presents simulations of the metastable helium
absorption lines in the transmission of WASP-107b. We employ
full 3D hydrodynamic simulations to model the dynamics of
evaporating planetary atmospheres, in which non-equilibrium
thermochemistry and ray-tracing radiative transfer are
co-evolved. The processes that populate and de-populate the
metastable state of neutral helium are included in the
thermochemical network and solved consistently. These allow
us to predict the mass loss rate, the temperature profile,
and synthetic observation in both \lya and \ext{He}; we note
that previous works often have to assume or prescribe the
first two.

By exploring the parameter space, we find a plausible model
for WASP-107b that also involves a stellar wind stronger
than the solar wind. This model launches a photoevaporative
outflow with a mass-loss rate of
$\dot{M}\simeq 1.0\times 10^{-9}~M_\oplus~\yr^{-1}$. The
predicted \ext{He} line profiles and light curves exhibit
reasonable agreement with existing
observations. Time-averaged transmission spectrum has a
$\sim 1:4 > 1:8$ line ratio, while the light curve holds a
considerable asymmetry about the transit median. A
comet-like tail trailing the planet is a natural explanation
of these observations. We argue that such a tail is the
result of a relatively strong stellar wind rather than
radiation pressure, because (1) the low overall abundance
of\ext{He}, and (2) the photon momenta deposited by
absorption and scattering are quickly thermalized by
collisional momentum relaxation. The incoming stellar wind
triggers shear instabilities, and the \ext{He} transit depth
fluctuates by $5-20~\%$ at a roughly $\sim 0.2~{\rm day}$
timescale consequently. Tuning the characteristics of
stellar wind in a plausible range does affect the spectral
line shapes through alternating the direction and
configuration of the tail, yet the equivalent width is
largely invariant. The \lya transmission spectra and light
curve synthesized using the fiducial Model 107-0 suggest
stronger light curve asymmetry and variabilities due to
shear instabilities. We propose that surveys of exoplanets
in \lya and \ext{He} simultaneously are crucial for
understanding planetary outflow and stellar wind.

Looking ahead, planetary outflow modulated by stellar
winds may extract positive or negative orbital angular
momentum from the planet via dynamical (anti-)friction
\citep{2020MNRAS.494.2327L}, and may then lead to migrations
of planetary orbits. Planetary magnetic fields and the
fields carried by stellar winds or coronal ejecta may also
play essential roles in shaping the dynamics of an
evaporating close-in planet. Future simulations should have
magnetohydrodynamics involved in these processes with
improved consistency, and potentially shed light on the
evolution of magnetic fields when compared to observations.

\vspace*{20pt}

This work is supported by the Center for Computational
Astrophysics of the Flatiron Institute, and the Division of
Geological and Planetary Sciences of the California
Institute of Technology. L. Wang acknowledges the computing
resources provided by the Simons Foundation and the San
Diego Supercomputer Center. We thank our colleagues
(alphabetical order): Philip Armitage, Zhuo Chen, Jeremy
Goodman, Xinyu Li, Mordecai Mac-Low, Songhu Wang, and Andrew
Youdin, for helpful discussions and comments.

\bibliography{planet_he.bib}
\bibliographystyle{aasjournal}
% \bibliographystyle{apj}

% END OF DOCUMENT
%
\end{document}